\RequirePackage{fix-cm}
\documentclass[twocolumn,epjc3]{svjour3} 
\smartqed  
\RequirePackage{graphicx}
\RequirePackage{xcolor}
\RequirePackage{amssymb}
\RequirePackage{esint}
\newcommand{\eqref}[1]{(\ref{#1})}
\journalname{Eur. Phys. J. C}
\def\Xint#1{\mathchoice
{\XXint\displaystyle\textstyle{#1}}
{\XXint\textstyle\scriptstyle{#1}}
{\XXint\scriptstyle\scriptscriptstyle{#1}}
{\XXint\scriptscriptstyle\scriptscriptstyle{#1}}
\!\int}
\def\XXint#1#2#3{{\setbox0=\hbox{$#1{#2#3}{\int}$}
\vcenter{\hbox{$#2#3$}}\kern-.5\wd0}}

\def\dashint{\Xint-}
\newcommand{\bqa}{\begin{eqnarray}}
\newcommand{\eqa}{\end{eqnarray}}
\newcommand{\nl}{\nonumber \\}
\newcommand{\mur}{\mu_{\scriptscriptstyle \rm R}}

\newcommand{\s}{~\,}
\def\eps{\epsilon}
\begin{document}

\title{Direct numerical evaluation of multi-loop integrals without contour deformation
}


\author{Roberto Pittau\thanksref{e1,addr1,addr2}
        \and
     Bryan Webber\thanksref{e2,addr3} 
}

\thankstext{e1}{e-mail: pittau@ugr.es}
\thankstext{e2}{e-mail: webber@hep.phy.cam.ac.uk}


\institute{Departamento de F\'isica Te\'orica y del Cosmos and CAFPE, Universidad de Granada, 18071 Granada, Spain\label{addr1}
           \and
           Theoretical Physics Department, CERN, 1211 Geneva 23, Switzerland\label{addr2}
           \and
University of Cambridge, Cavendish Laboratory, J.J. Thomson Avenue, Cambridge, UK\label{addr3}
}

\date{Received: date / Accepted: date}

\maketitle

\begin{abstract}
We propose a method for computing numerically integrals defined via $i \epsilon$ deformations acting on single-pole singularities. We achieve this without an explicit analytic contour deformation. 
Our solution is then used to produce precise Monte Carlo estimates of multi-scale multi-loop integrals directly in Minkowski space. We corroborate the validity of our strategy by presenting several examples ranging from one to three loops. When used in connection with four-dimensional regularization techniques, our treatment can be extended to ultraviolet and infrared divergent integrals. 

\end{abstract}
\section{Introduction}
\label{sec:intro}
%
%
The ever-increasing precision of data from particle physics experiments requires a comparable or better level of precision in theoretical predictions, both to establish the parameters of the Standard Model and to search for physics beyond it.  To achieve such precision requires the computation of multi-loop amplitudes. 
A fundamental ingredient of such calculations is the evaluation of master loop integrals (MIs), in terms of which the problem is reduced. This can be performed by analytic, semi-numerical  or fully numerical techniques (see \cite{Heinrich:2020ybq} for a recent review).  Analytic methods are very successful when the class of functions that contribute to the result is known, which usually happens when the number of internal and external masses is limited. However, such a-priori knowledge is not always available, especially when the number of scales increases, so that in these cases one would like to be able to compute MIs numerically, for instance by Monte Carlo (MC) techniques.

In the numerical computation of MIs, an important problem is the appearance of integrable threshold singularities, where single poles are moved away from the real integration domain by the $i \epsilon$ prescription.
These singularities require special treatment, such as a contour deformation into the complex plane \cite{Soper:1999xk,Binoth:2000ps,Binoth:2005ff,Nagy:2006xy}, or vanishing-width extrapolations methods \cite{deDoncker:2004fb,Yuasa:2011ff,deDoncker:2017gnb,Baglio:2020ini}. Contour deformations are usually controlled by some parameter 
whose value should be not too small, to guarantee numerical accuracy, and not too large, to avoid crossing branch cuts. In extrapolation methods a series of integrals should be determined that converges to the right value while keeping the computation time low.

This paper explains how integrals defined through the $i \epsilon$ prescription acting on first-order poles can be evaluated numerically without deforming the integration contour into the complex plane, and how this can be employed to compute MIs appearing in multi-loop calculations. In addition, we demonstrate that this strategy {allows one} to compute recursively higher-loop functions in terms of lower-loop ones. {Other semi-numerical methods relying on one-loop-like objects to build higher loops can be found in \cite{Ghinculov:1996vd,Guillet:2019hfo,Bauberger:2019heh}.
In \cite{Ghinculov:1996vd} a Wick rotation of the loop momentum is needed to avoid singularities. The Feynman parameter space is used in \cite{Guillet:2019hfo}, and a contour deformation in \cite{Bauberger:2019heh}. Our method works directly in Minkowski space and avoids contour deformations.}

The  structure  of  the  paper  is  as  follow.  Sect.~\ref{sec:2} details our approach. In Sect.~\ref{sec:3} we use it to integrate numerically threshold singularities after an analytic integration over the energy-components of the loop momenta. Sect.~\ref{sec:4} explains how to {\em glue} together lower-loop structures to compute numerically certain classes of higher-loop MIs. Finally, in Sects.~\ref{sec:5} and \ref{sec:6} we extend our treatment to ultraviolet and infrared divergent configurations regularized via the four-dimensional method of \cite{Pittau:2012zd}.

\section{Avoiding contour deformation}
\label{sec:2}
In this section we present two methods which avoid contour deformation.
The first method uses complex analysis, while the second approach directly works with the original integrand. The two procedures are equivalent, in that they give rise to the same mappings. The numerical results presented in the paper are obtained with method 1 and cross-checked with method 2. 

\subsection{Method 1}
\label{sec:method1}
For the sake of clarity, distinct letters (with or without additional subscripts) are used to denote variables ranging in different intervals. In particular, we employ
$x$ when $-1 \le x \le 1$, $y$ if $0 \le y \le 1$,
$\sigma$ provided $-\infty < \sigma < \infty$.
Finally, $0 \le \rho \le 1$ stands for a random  Monte Carlo (MC) variable.

The core of the procedure is a change of variable such that the $1/(x+i\epsilon)$ behaviour of the integral
\bqa
\label{eq:intIa}
I := \lim_{\epsilon \to 0 }\int_{-1}^1 dx \frac{1}{x+i\epsilon}
\eqa
is flattened with $x \in \mathbb{R}$.
This is obtained by imposing
\bqa
\label{eq:chvarx}
x+i\epsilon= e^{i \pi(1-z)},
\eqa
where $z$ is a new complex integration variable. In fact, inserting \eqref{eq:chvarx} in \eqref{eq:intIa} gives the desired result,
\bqa
\label{eq:intIb}
I = -i \pi \landupint_0^1 d z.
\eqa
Eq.~\eqref{eq:intIb} evaluates to $-i \pi$ along any curve in the $z$ complex plane connecting $z= 0$ to $z= 1$ when $\epsilon \to 0$. We use this freedom to impose $x \in \mathbb{R}$ by parametrizing
\bqa
\label{eq:z}
z= \alpha + i \beta \s {\rm with} \s \alpha,\beta \in \mathbb{R} \s {\rm and } \s \frac{\epsilon}{\pi} \le \alpha \le 1-\frac{\epsilon}{\pi}.
\eqa
Inserting \eqref{eq:z} in \eqref{eq:chvarx} gives
\bqa
x= e^{\pi \beta} \cos[\pi (1-\alpha)]
+ i \{e^{\pi \beta} \sin[\pi (1-\alpha)]-\epsilon \},
\eqa
which is real when 
$\displaystyle
\pi \beta= \ln \frac{\epsilon}{\sin[\pi (1-\alpha)]}$,
namely
\bqa
x= x_\alpha := \frac{\epsilon}{\tan[\pi (1-\alpha)]}.
\eqa
Therefore 
\bqa
\label{eq:par1}
dz= d\alpha \left(1+i \frac{d \beta}{d \alpha}\right),
\eqa
which gives
\bqa
\label{eq:intz01}
\landupint_0^1 d z = \lim_{\epsilon \to 0} \frac{1}{g_\epsilon} \int_{{\epsilon}/{\pi}}^{1-{\epsilon}/{\pi}}\!\!\!\!d\alpha 
\left( 1+i\frac{x_\alpha}{\epsilon} \right),
\s g_\epsilon := 1-\frac{2\epsilon}{\pi},
\eqa
where $g_\epsilon$ has been introduced to impose the normalization to 1 also for small but not vanishing values of $\epsilon$.
{In summary, after changing variable as in \eqref{eq:chvarx}, the requirement $x \in \mathbb{R}$ determines the relation between 
${\Re}e(z)$ and ${\Im}m(z)$.}

Armed with these results, we generalize \eqref{eq:intIa} to an integration over a function
\bqa
\label{eq:f}
f(x)= \phi(x)/(x+i \epsilon),
\eqa
with $\phi(x)$ sufficiently smooth at $x= 0$,
\footnote{From now on, we omit $\lim_{\epsilon \to 0}$ and consider $\epsilon$ as an infinitesimal parameter.}
\bqa
\label{eq:Ifa}
I_f := \int_{-1}^{1} dx\, f(x)= \int_{0}^{1} dy\,
\big[f(-y)+f(y)\big].
\eqa
Splitting the integration region of \eqref{eq:intz01} into the two sectors with $x_\alpha < 0$ or $x_\alpha > 0$ gives
\bqa
\label{eq:Ifb}
I_f &=& -\frac{i \pi}{g_\epsilon}
\int_{{\epsilon}/{\pi}}^{1/2}\!\! d\alpha \nl
&&\times\left[
 \left(1-i\frac{y_\alpha}{\epsilon}\right)\phi(-y_\alpha)
+\left(1+i\frac{y_\alpha}{\epsilon}\right)\phi(y_\alpha)
\right],
\eqa
where $y_\alpha:= {\epsilon}/{\tan(\alpha \pi)}$.
Eq.~\eqref{eq:Ifb} can be translated to a MC language  by looking for the local density $g(y)$ that corresponds to a change of variable $d \rho= g(y) dy$ reabsorbing the singular behaviour of the integrand of \eqref{eq:Ifa},
\bqa
\label{eq:Ifc}
I_f= \int_0^1 d \rho\, \frac{f(-y)+f(y)}{g(y)},
\eqa
with $\int_0^1 dy\, g(y)=1$. By 
comparing \eqref{eq:Ifc} to \eqref{eq:Ifb} one determines
\bqa
\label{eq:mapr}
g(y)= \frac{2 \epsilon}{\pi (y^2+\epsilon^2)},\s
y = \frac{\epsilon}{\tan(\alpha \pi)},\s \alpha = \frac{\epsilon}{\pi}+\frac{\rho g_\epsilon}{2}.
\eqa

The mapping of \eqref{eq:mapr} optimizes
the integration over the real part of $z$ (see  \eqref{eq:par1}). This gives stable numerical results when $\phi(x)$
is such that the $y_\alpha/\epsilon$ terms in \eqref{eq:Ifb} are suppressed. When this is not the case, they generate a large contribution to the variance and, in order to flatten them, the parametrization complementary to \eqref{eq:par1} is necessary,
\bqa
\displaystyle dz= d\beta \left(\frac{d \alpha}{d \beta}+i\right),
\eqa
which gives
\bqa
\label{eq:Ifd}
I_f&=& - \frac{i \pi}{g_\epsilon} \int_{\beta_-}^{\beta_+}
d\beta \nl
&& \times \left[
 \left( \frac{\epsilon}{-y_\beta} +i\right) \phi(-y_\beta)
-\left( \frac{\epsilon}{ y_\beta} +i \right)\phi( y_\beta)
\right],
\eqa
where
\bqa
y_\beta:= e^{\pi\beta}\sqrt{1-\left(\frac{\epsilon}{e^{\pi \beta}}\right)^2},\s\beta_-= \frac{1}{\pi}\ln \frac{\epsilon}{\sin \epsilon},\s\beta_+= \frac{\ln \epsilon}{\pi}.  \nonumber
\eqa
Again, \eqref{eq:Ifd} is correctly normalized also for a small but not vanishing $\epsilon$.
Comparing \eqref{eq:Ifc} to \eqref{eq:Ifd} gives now
\bqa
\label{eq:mapi}
g(y)&=& -\frac{g_\epsilon}{\ln (\sin \epsilon)}\frac{y}{ (y^2+\epsilon^2)},\s
y = e^{\pi\beta}\sqrt{1-\left(\frac{\epsilon}{e^{\pi \beta}}\right)^2},\nl
\beta&=& \frac{\ln(\epsilon)-\rho \ln(\sin \epsilon)}{\pi}.
\eqa
\paragraph{Multichanneling}
$~$ \vskip 5pt
\noindent Flattening the whole $1/(x+i\epsilon)$ behaviour of \eqref{eq:f} requires a merging of \eqref{eq:mapr} and \eqref{eq:mapi}, whose densities we dub $g_{1} (y)$ and $g_{2}(y)$.  This can be achieved via a
multichannel approach with combined density \\
$g_{\rm c}(y) := \alpha_1 g_{1}(y)+\alpha_2 g_{2}(y)$ and $ \alpha_1+\alpha_2= 1$,
\bqa
\label{eq:mca}
I_f = \int_0^1 d \rho\, \frac{f(-y)+f(y)}{g_{\rm c}(y)}.
\eqa
In \eqref{eq:mca}, $\rho$ is generated according to  the distribution $g_{1,2}(y)$  with probability $\alpha_{1,2}$, and
the {\it a-priori} weights $\alpha_{1,2}$ can be optimized as described in \cite{Kleiss:1994qy}.

To reduce the variance when $\phi(x)$ peaks inside $-1 \le x \le  1$ in a known way, it is also possible to include an arbitrary numbers of further channels $g_i(y)$ $(i > 2)$. However, care must be taken due to the fact that the MC weight of \eqref{eq:Ifc} includes both the $f(-y)$ and $f(y)$ contributions. To determine the corresponding density we observe that
\bqa
\label{eq:normgi}
\int_0^1 dy \,\big[g_i(-y)+g_i(y)\big] = \int_{-1}^{1} dx\, g_i(x),
\eqa
which means that if $x$ is randomly chosen
in $-1 \le x \le 1$, the density is $g_i(-|x|)+ g_i(|x|)$. Hence, the MC weight is $\big(f(-y)+f(y)\big)/\big(g_i(-y)+g_i(y)\big)$, with $g_i$
normalized such that
\bqa
\int_{-1}^{1} dx\, g_i(x)= 1.
\eqa
In summary, with $N_{\rm ch}$ channels (including $g_1$ and $g_2$), the more general multichannel MC mapping reads
\bqa
\label{eq:mcb}
I_f = \int_0^1 d \rho\, \frac{f(-y)+f(y)}{g_{\rm tot}(y)},\s
d \rho =  g_{\rm tot}(y) dy,
\eqa
where
\bqa
g_{\rm tot}(y) &=&g_{\rm c}(y) +
\sum_{i=3}^{N_{\rm ch}} \alpha_i (g_i(-y)+g_i(y)),
\eqa
with arbitrary (but self-adjustable) weights fulfilling
$\sum_{i=1}^{N_{\rm ch}} \alpha_i= 1$.
In the actual MC used to produce the results presented in this paper we superimpose on $g_{\rm c}(y)$ a flat distribution $g_3(x)= 1/2$ 
and a channel 
\bqa
g_4(x)= \frac{1}{2 \ln\frac{1+\delta}{\delta}} \frac{1}{1-|x|+\delta},~~\delta= 10^{-4},
\eqa
which takes care of peaks around $|x|= 1$.
\paragraph{Principal value integrals}
$~$ \vskip 5pt
\noindent It is often useful to deal with improper integrals, whose behaviour at large values of the integration variables is defined via the Cauchy principal value. The fact that the two symmetric points with respect to $x=0$ are always considered together, makes the use of \eqref{eq:mcb} very convenient. 
As a matter of notation, we define
\bqa
\dashint  d \sigma := \lim_{\Lambda \to \infty} \int_{-\Lambda}^\Lambda d \sigma,
\eqa
which can be mapped onto the interval $[-1,1]$ by changing variable,
\bqa
\sigma= \frac{x}{1-x^2}.
\eqa
Thus, for instance,
\bqa
\label{eq:cauchint}
\dashint d \sigma  \frac{\phi(\sigma)}{\sigma + i \epsilon}=
 \int_{-1}^{1} dx\,\frac{1+x^2}{1-x^2} \phi \Big(\frac{x}{1-x^2}\Big)\frac{1}{x+ i \epsilon},
\eqa
where we understand the symmetric treatment of \eqref{eq:mcb}, so that \eqref{eq:cauchint} is well defined even when $\phi(\sigma)$ approaches a constant as
$\sigma \to \pm \infty$.
\paragraph{Multiple integrals}
$~$ \vskip 5pt
\noindent Eq.~\eqref{eq:mcb} can be easily extended to $n$-fold integrals of the type
\bqa
I_{f,n} := \int_{-1}^1 \prod_{j=1}^{n} \big( dx_j \big) 
f( \{ x\}), 
\eqa
with
$
f( \{ x\})= \phi(\{ x\})/{\prod_{j=1}^{n}(x_j+i \epsilon)}$.
Our notation is such that
$\{ x\}= x_1, x_2, \cdots, x_n$ and $\phi(\{ x\})$ is a smooth function at $\{ x\}= \{ 0\}$. The result is
\bqa
\label{eq:mcn}
I_{f,n} = \int_0^1  \prod_{j=1}^{n} \big( d\rho_j \big)
\frac{f(-\{ y\})+f(\{ y\})}{\prod_{j=1}^{n} g_{\rm tot}(y_j)},
\eqa
where $\s d \rho_j= g_{\rm tot}(y_j) dy_j$ and
the numerator stands for a sum over the $2^n$ terms with positive or negative arguments. For instance, when $\{ y\}= y_1,y_2$,
\bqa
&&f(-\{ y\})+f(\{ y\})= \\
&&~~f(-y_1,-y_2)+f(-y_1,y_2)+f(y_1,-y_2)+f(y_1,y_2).
\nonumber
\eqa
{Equation \eqref{eq:mcn} can be generalized to more poles per variable, moved away from arbitrary domains $\in \mathbb{R}$, either by partial fractioning the integrand or by splitting the integration region into sub-intervals. However, configurations like that never appear in what follows, so we do not pursue a detailed analysis in this direction.}

\subsection{Method 2}
\label{sec:method2}
As an alternative to the above method, one can apply separate changes of variables to flatten the real and imaginary parts of the pole factor(s) in the integrand.  Consider the integral
\bqa
\label{eq:meth2}
{\cal I}_1[\phi] := \int_{-a}^a dx\frac{\phi(x)}{x+i\eps} =\int_0^a
dx\frac{x\phi^-(x)-i\eps \phi^+(x)}{x^2+\eps^2},
\eqa
where $\phi^\pm(x)=\phi(x)\pm \phi(-x)$.  We can write this as
 \bqa\label{eq:rtheta}
{\cal I}_1[\phi] &=& \int_0^{a/\eps} dy\frac{y\phi^-(\eps y)-i\phi^+(\eps
  y)}{1+y^2}\nl
&=& \int_0^{r_m} dr\,\phi^-(\eps\sqrt{{\rm e}^{2r}-1})
-i\int_0^{\theta_m}d\theta\,\phi^+(\eps\tan\theta),\nl
\eqa
where
\bqa\label{eq:rmdef}
r_m :=\ln(1+a^2/\eps^2)/2\,,\;\;\theta_m := \arctan(a/\eps).
\eqa
Each of these integrals has optimal variance reduction (in the absence of information about $\phi$) and is therefore suited to numerical integration as long as $\phi(x)$ is smooth at $x = 0$.
Note that the $\eps/(x^2+\eps^2)$ and $x/(x^2+\eps^2)$ behaviours of \eqref{eq:meth2} correspond to the local densities in \eqref{eq:mapr} and \eqref{eq:mapi}, respectively.

The method is easily generalised to two variables. Consider
\bqa\label{eq:2var}
{\cal I}_2[\phi] &:=& \int_{-a}^a dx_1
dx_2\frac{\phi(x_1,x_2)}{(x_1+i\eps) (x_2+i\eps)}\nl
&=& \int_0^a dx_1 dx_2\frac{\Phi(x_1,x_2;\eps)}{(x_1^2+\eps^2)
  (x_2^2+\eps^2)},
\eqa
where
\bqa
\Phi(x_1,x_2;\eps)=x_1x_2\phi^{11}-i\eps(x_1\phi^{10}+x_2\phi^{01}) -\eps^2\phi^{00}
\eqa
with
\bqa\label{eq:fij}
&&\phi^{00} = \nl
&&~~\phi(x_1,x_2) +\phi(-x_1,x_2) +\phi(x_1,-x_2) +\phi(-x_1,-x_2),\nl 
&&\phi^{10} = \nl 
&&~~\phi(x_1,x_2) -\phi(-x_1,x_2) +\phi(x_1,-x_2) -\phi(-x_1,-x_2),\nl 
&&\phi^{01} = \nl 
&&~~\phi(x_1,x_2) +\phi(-x_1,x_2) -\phi(x_1,-x_2) -\phi(-x_1,-x_2),\nl 
&&\phi^{11} = \nl 
&&~~\phi(x_1,x_2) -\phi(-x_1,x_2) -\phi(x_1,-x_2) +\phi(-x_1,-x_2).\nl
\eqa
For MC evaluation, we proceed as follows:  for each shot, generate
$x_{1r}, x_{1t}, x_{2r}, x_{2t}$ where
\bqa\label{eq:xrxt}
&&x_{1r}=\eps\sqrt{{\rm e}^{2r_1}-1},\;\;\; 
x_{1t}= \eps\tan\theta_1,\nl
&&x_{2r}=\eps\sqrt{{\rm e}^{2r_2}-1},\;\;\; 
x_{2t}= \eps\tan\theta_2
\eqa
where
\bqa
0<r_{1,2}<r_m,\;\;\;
0<\theta_{1,2}<\theta_m
\eqa
uniformly, with $r_m$ and $\theta_m$ as in (\ref{eq:rmdef}).
In (\ref{eq:fij}), set $x_1= x_{1t}$ when the first
superscript is 0 and  $x_1= x_{1r}$ when it is 1,  and similarly for
$x_2$ according to the second superscript.  The weights for the real
and imaginary parts are then
\bqa\label{eq:wrwt}
w_r = \phi^{11}r_m^2-\phi^{00}\theta_m^2,\;\;\;
w_i =-(\phi^{10}+\phi^{01})r_m\theta_m.
\eqa

The generalisation of (\ref{eq:2var}) to $n$ variables is clear:
$\phi^{\{k_j\}}$ has superscript $k_j=1$ in the $j$th location when there is an $x_j$ in the integrand, otherwise $k_j=0$.  The symmetrized function $\Phi$ becomes
\bqa
\Phi(\{x_j\};\eps) =
(-i\eps)^n\sum_{\{k_j=0,1\}}\prod_{j=1}^n(ix_j/\eps)^{k_j}\phi^{\{k_j\}}(\{x_j\}),\nl
\eqa
where
\bqa\label{eq:fkj}
\phi^{\{k_j\}}(\{x_j\})= \sum_{\{l_j=0,1\}} \prod_{j=1}^n (-1)^{k_j l_j} \phi\left(\{(-1)^{l_j}x_j\}\right).
\eqa

For MC evaluation, for each shot, generate two
points in the $n$-dimensional hypercube
\bqa\label{eq:xjrt}
x_{jr}=\eps\sqrt{{\rm e}^{2r_j}-1},\;\;\;
x_{jt}= \eps\tan\theta_j,
\eqa
where again $0<r_j<r_m$ and $0<\theta_j<\theta_m$
uniformly.  In (\ref{eq:fkj}), set $x_j= x_{jt}$ when $k_j=0$
and  $x_j= x_{jr}$ when $k_j=1$.  The weight is then
\bqa
w_r+i w_i =
(-i\theta_m)^n\!\!\!\!\sum_{\{k_j=0,1\}}\prod_{j=1}^n (ir_m/\theta_m)^{k_j}\phi^{\{k_j\}}(\{x_j\}).\nl
\eqa
Note that each shot involves $2n$ random numbers for  $\{x_{jt}\}$
and  $\{x_{jr}\}$, and then $4^n$ function evaluations at $x_j=\pm
x_{jt}$ and $\pm x_{jr}$, so the computation time increases rapidly with the number of variables.

\subsection{Choosing $\epsilon$}
\label{sec:seteps}
\begin{figure}
\vskip -4.3cm
\hskip -3.1cm
\includegraphics[width=5.7in]{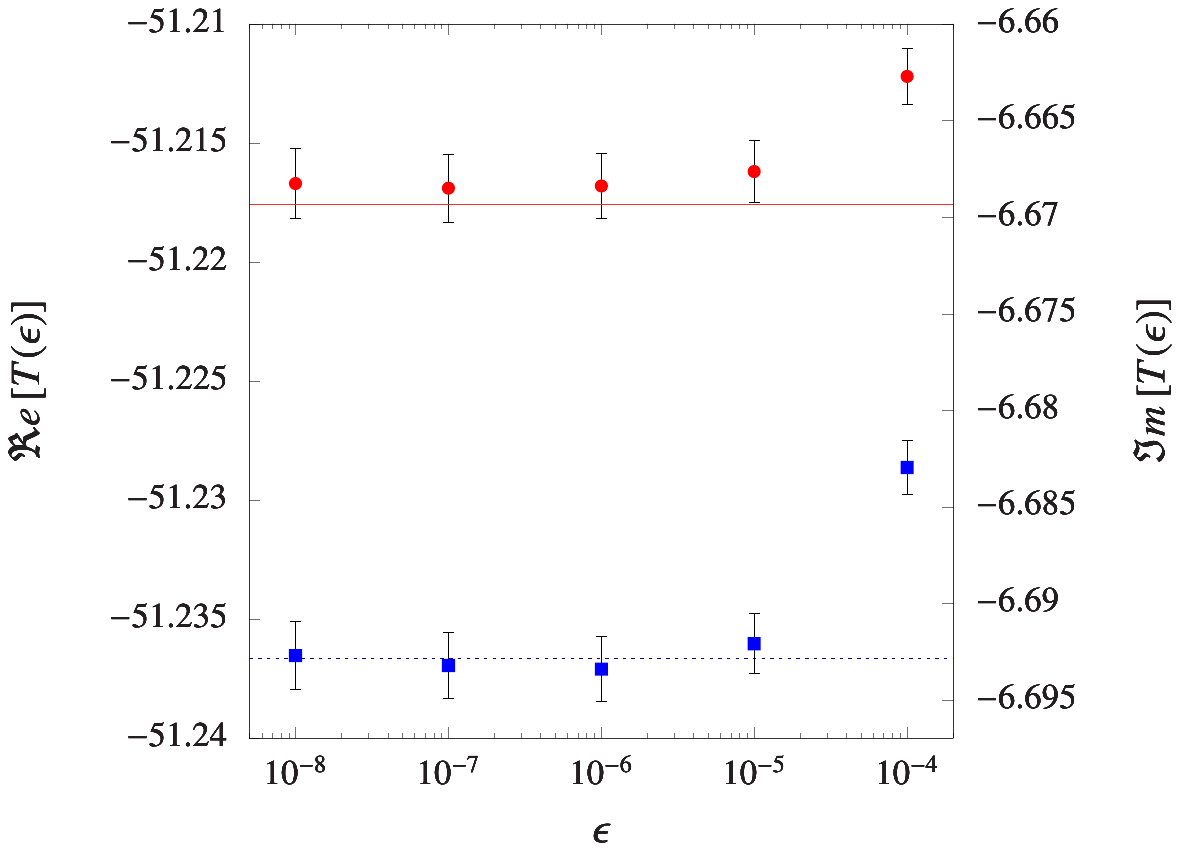}
\vskip -10cm
\caption{MC results for the real (red bullets) and imaginary (blue squares) part of \eqref{eq:TepsMC}. They have been obtained with $10^{10}$ MC shots per point, corresponding to $8 \times 10^{10}$ calls to the integrand. To minimize the statistical fluctuations, the same sequence of random numbers is used for all values of $\eps$.}
\label{fig:eps}   
\end{figure}

Here we perform a study of the value of $\eps$ to be used in practice. More specifically, we compare the numerical and analytic determinations of the three-fold test integral
\bqa
\label{eq:TepsMC}
T (\epsilon) &=& 
\int_{-1}^{1} \prod_{j=1}^{3} \left( \frac{dx_j}{x_j+i \epsilon} \right) 
\sum_{j_1,j_2,j_3= 0}^{1}
x_1^{j_1} x_2^{j_2} x_3^{j_3} \\
\label{eq:Tepsana}
&=& (8-6\pi^2)+ i\pi (\pi^2-12),
\eqa
whose behaviour at $x_j \sim 0$ mimics a typical multi-dimensional environment.  
The result of this comparison is given in Fig.~\ref{fig:eps}, where the solid (dashed) line represents the real (imaginary) part of \eqref{eq:Tepsana}.
Bullets and squares with errors are the MC predictions for ${\Re}e\left[T(\eps)\right]$ and ${\Im}m\left[T(\eps)\right]$, respectively.
{To quantify the effect of a nonzero $\epsilon$ on the MC estimate $Q_{\rm MC}(\epsilon)\pm \Delta Q(\epsilon)$ of a known quantity $Q$, it is convenient to introduce the estimators
\bqa
\Delta_1(\epsilon) = \frac{|Q-Q_{\rm MC}(\epsilon)|}{|Q|},\,
\Delta_2(\epsilon) = \max\bigg(\!\epsilon,\frac{\Delta Q(\epsilon)}{|Q|} \bigg). 
\eqa
Requiring the $\epsilon \ne 0$ bias on $Q_{\rm MC}(\epsilon)$ to be of the order of the maximum between $\epsilon$ and the relative MC error gives the condition
\bqa
\label{eq:ratio}
R(\epsilon,Q) := \frac{\Delta_1(\epsilon)}{\Delta_2(\epsilon)} \sim {\cal O}(1).
\eqa
Table \ref{tab:eps} reports the $R$ value of the entries of Fig.~\ref{fig:eps} and their MC accuracy defined as
\bqa
\label{eq:mcacc}
\delta(\epsilon) := \frac{\max\big(\Delta {\Re}e[T(\epsilon)],\Delta {\Im}m[T(\epsilon)]\big)}{|T_{\rm MC}(\epsilon)|}.
\eqa}
From Fig.~\ref{fig:eps} {and Table \ref{tab:eps}} we infer that a range $10^{-8} \leq \eps \leq 10^{-6}$ is adequate to achieve MC estimates accurate at the level of {three} parts in $10^5$. 
{Since the results presented in this paper are never more accurate than this, we set, for definiteness, $\eps= 10^{-7}$.} 
{However, the last row of Table \ref{tab:eps} shows that numerically stable predictions are produced also with a smaller $\epsilon$ and a larger MC statistics. From this, we deduce that the $\epsilon \ne 0$ bias can be 
reduced to be negligible in most practical applications, and that the accuracy of our method is driven by the MC error.
}
\begin{table}
\caption{{The ratio in \eqref{eq:ratio} for the real and imaginary parts of $T(\epsilon)$ in \eqref{eq:TepsMC} computed with $10^{10}$ (3$\times 10^{11}$) MC shots
when $\epsilon \ge 10^{-8}$ ($\epsilon = 10^{-12}$). The last column reports the MC accuracy defined in \eqref{eq:mcacc}.}}
\label{tab:eps}       
\begin{tabular}{lccc}
\hline\noalign{\smallskip}
 {$\epsilon$}   & {$R(\epsilon,{\Re}e[T])$} & {$R(\epsilon,{\Im}m[T])$} & {$\delta(\epsilon)$}\\
\noalign{\smallskip}\hline\noalign{\smallskip}
    {$10^{-4}$} & {1.1}& {7.0} 
    &{3$\times 10^{-5}$}\\
    {$10^{-5}$} & {1.1}& {0.5}
    &{3$\times 10^{-5}$}\\
    {$10^{-6}$} & {0.6}& {0.3}
    &{3$\times 10^{-5}$}\\ 
    {$10^{-7}$} & {0.5}& {0.2}
    &{3$\times 10^{-5}$}\\
    {$10^{-8}$} & {0.6}& {0.1}
    &{3$\times 10^{-5}$}\\ 
    {$10^{-12}$}& {0.02} &  {1.5}
    &{7$\times 10^{-6}$} \\
\noalign{\smallskip}\hline
\end{tabular}
\end{table}

\section{A semi-numerical integration algorithm for MIs}
\label{sec:3}
In this section we illustrate how the approach of Sect.~\ref{sec:2} can be successfully applied to produce stable and precise semi-numerical MC estimates of loop MIs. This is achieved in two steps.
Firstly, we integrate analytically over the energy components of the loop momenta, which is always doable by means of the Cauchy integral theorem. In addition, depending on the case at hand, some of the loop angular integrals can also be performed analytically. In this way, integral representations of MIs can be easily obtained. 
Secondly, we give up any attempt towards a fully analytic integration, which may be difficult, and integrate numerically over the left-over loop components. The integrand to be evaluated is usually plagued by threshold singularities. Single poles migrate towards the real integration domain for some kinematic configurations, so that a blind numerical integration over denominators deformed by the Feynman $i \epsilon$ prescription gives large errors. However, this is precisely the situation for which our approach is designed. We mitigate these problems by retaining a finite small value of $\epsilon$, flattening the real and imaginary parts of pole contributions, and applying multichannel mappings.~\footnote{The use of threshold counterterms~\cite{Kermanschah:2021wbk} could further improve the precision of our approach.} In what follows we illustrate the performance of this strategy by means of two examples.
\subsection{A one-loop example}
\label{sec:3a}
\begin{figure}
\vskip -4.2cm
\hskip -4.7cm
\includegraphics[width=6.5in]{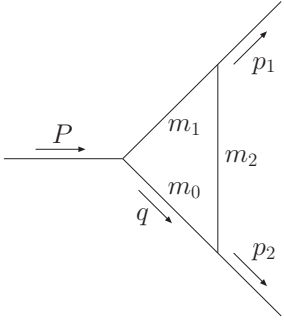}
\vskip -3.9cm
\caption{The scalar three-point one-loop function with arbitrary kinematics and  masses $C(P^2,p^2_1,p^2_2,m_0,m_1,m_2)$.}
\label{fig:1}   
\end{figure}
Consider the three-point function of 
Fig.~\ref{fig:1} in the case $m_0=m_1=m_2= m$, $p^2_1=p^2_2=0$ and timelike $P$. Rescaling all momenta by $m$
\bqa
\label{eq:omegaq}
&&\frac{q}{m} = 
(t,\rho c_\theta,\rho s_\theta
s_\phi,\rho s_\theta c_\phi),~\frac{P}{m} = (\sqrt{\tau},0,0,0), \\
&&\frac{p_1}{m} = \frac{\sqrt{\tau}}{2} (1,1,0,0),~ 
\frac{p_2}{m} = \frac{\sqrt{\tau}}{2} (1,-1,0,0),
\eqa
gives
\bqa
\label{eq:eqC0III}
C_0&:=& C(P^2,0,0,m,m,m)= \frac{2 \pi}{m^2} \int_{-1}^1 d c_\theta \int_0^\infty d \rho \rho^2 \nl
&& \times \int_{-\infty}^{\infty} dt \frac{1}{(\sigma_0+i \epsilon)(\sigma_1+i \epsilon)(\sigma_2+i \epsilon)},
\eqa
with
\bqa
\sigma_0= \frac{q^2}{m^2}-1,~
\sigma_1= \frac{(q-P)^2}{m^2}-1,~
\sigma_2= \frac{(q-p_2)^2}{m^2}-1. \nonumber
\eqa
One splits
\bqa
\frac{1}{\sigma_2+i \epsilon}&=&
\frac{1}{2 R_2}
\Biggl( 
 \frac{1}{t- \sqrt{\tau}/2-R_2+i \epsilon} \nl 
&&-\frac{1}{t- \sqrt{\tau}/2+R_2-i \epsilon}
\Biggr)
\eqa
with $\displaystyle R^2_2 := \rho^2+ \frac{\tau}{4}+\sqrt{\tau}\rho c_\theta+1$.
Thus
\bqa
\int_{-1}^1 d c_\theta
\frac{1}{\sigma_2+i \epsilon} &=&
\frac{1}{\sqrt{\tau} \rho} \Bigl(
 \ln \frac{t-\sqrt{\tau}/2-R^-_2 + i \epsilon}{t-\sqrt{\tau}/2-R^+_2 + i \epsilon} \nl
 && +\ln \frac{t-\sqrt{\tau}/2+R^-_2 - i \epsilon}{t-\sqrt{\tau}/2+R^+_2 - i \epsilon}
\Bigr),
\eqa
where $\displaystyle R^\pm_2 := \sqrt{({\sqrt{\tau}}/{2}\pm \rho)^2+1}$. The cut of the logarithms with $+ i \epsilon$
($- i \epsilon$) is in the lower (upper) $t$ complex half-plane, so that the integration over $t$ in
\eqref{eq:eqC0III} is trivial once one rewrites
\bqa
&&\!\!\frac{1}{(\sigma_0+i \epsilon)(\sigma_1+i \epsilon)} =
\frac{1}{4 R^2_0}
\Bigl(
 \frac{1}{t-R_0+i \epsilon}
-\frac{1}{t+R_0-i \epsilon}
\Bigr) \nl 
&&~\times \Bigl(
 \frac{1}{t-\sqrt{\tau}-R_0+i \epsilon}
-\frac{1}{t-\sqrt{\tau}+R_0-i \epsilon}
\Bigr), \nonumber  
\eqa
with $\displaystyle R^2_0 := \rho^2+1$. The results is
\bqa
\label{eq:c0III}
C_0 &=& \frac{2 i \pi^2}{m^2 \tau}
\int_{-\infty}^{\infty}\! \frac{dr}{r-i \epsilon}
\Big[
 \theta\bigl(r+{\textstyle \frac{\sqrt{\tau}}{2}}-1\bigr)
 L\bigl(r+{\textstyle \frac{\sqrt{\tau}}{2}},\sqrt{\tau}\bigr) \nl
&&+\theta(r-{\textstyle \frac{\sqrt{\tau}}{2}}-1) L(r-\textstyle{\frac{\sqrt{\tau}}{2}},-\sqrt{\tau})
\Big],
\eqa
where
\bqa
L(r,\sqrt{\tau}) :=
\ln \frac{r-\frac{\sqrt{\tau}}{2}+R(r,-\sqrt{\tau})- i \epsilon}{r-\frac{\sqrt{\tau}}{2}+R(r,\sqrt{\tau}) - i \epsilon}
\nonumber
\eqa
and
\bqa 
R(r,\sqrt{\tau}) := \sqrt{{\tau}/{4}+ r^2+ \sqrt{\tau} \sqrt{r^2-1}}.
\nonumber 
\eqa
When $\sqrt{\tau} > 2$, the first integrand of \eqref{eq:c0III} develops a pole at $r= i \epsilon$ that migrates towards the integration region in the limit $\epsilon \to 0$. Treating this  with the strategy of Sect.~\ref{sec:2} gives the results presented in Table \ref{tab:5c}.

\begin{table*}
\caption{Numerical estimates of the one-loop integral \eqref{eq:c0III} multiplied by $m^2$, compared to the analytic result of \cite{vanHameren:2010cp}.  Numbers obtained with $2 \times  10^7$ MC points. The MC errors are indicated between parentheses.}
\label{tab:5c}       
\begin{tabular}{rll}
\hline\noalign{\smallskip}
 $\sqrt{\tau}$   & MC result & Analytic result \\
\noalign{\smallskip}\hline\noalign{\smallskip}
    0.01     & 9.85(2)$\times 10^{-7}$ $-$$i$  4.9408(92)   & 0  $-$$i$  4.9348 \\ 
    0.2      & 9.90(1)$\times 10^{-7}$ $-$$i$  4.9482(55)   & 0  $-$$i$  4.9513 \\  
    0.5      & 1.0233(8)$\times 10^{-6}$ $-$$i$  5.0361(41)   & 0  $-$$i$  5.0412 \\  
    1.99     & 1.4341(9)$\times 10^{-5}$ $-$$i$  1.0783(3)$\times 10^{1}$ & 0  $-$$i$  1.0782$\times 10^{1}$\\  
    2.01     & 1.5350(6)    $-$$i$  1.2006(3)$\times 10^{1}$ & 1.5343  $-$$i$ 1.2006$\times 10^{1}$\\  
    10       & 1.4216(3)    +$i$  5.5007(30)$\times 10^{-1}$ & 1.4216  +$i$ 5.5030$\times 10^{-1}$ \\  
    10$^2$   & 2.8562(8)$\times 10^{-2}$ +$i$  3.6999(15)$\times 10^{-2}$       & 2.8557$\times 10^{-2}$  +$i$ 3.6990$\times 10^{-2}$ \\  
    10$^4$   &   5.7141(20)$\times 10^{-6}$ +$i$ 1.6258(5)$\times 10^{-5}$ &  5.7116$\times 10^{-6}$ +$i$ 1.6258$\times 10^{-5}$ \\
\noalign{\smallskip}\hline
\end{tabular}
\end{table*}

\subsection{A two-loop example}
\label{sec:3b}
\begin{figure}
\vskip -4.5cm
\hskip -5.4cm
\includegraphics[width=6.5in]{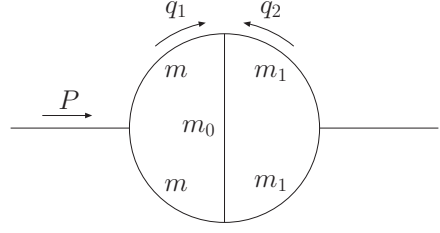}
\vskip -4.7cm
\caption{The two-loop self-energy diagram $S_2(m,m_0,m_1)$.}
\label{fig:1i}   
\end{figure}

We study the two-loop self-energy scalar diagram of Fig.~\ref{fig:1i}.
For a timelike  
$P/m= (\sqrt{\tau},{\vec 0})$ and $m_1= m$ it reads
\bqa
S_2 := S_2(m,m_0,m) = \frac{1}{m^2} \!\int  d^4 \omega_1   d^4 \omega_2 \prod_{j=1}^5 \frac{1}{\sigma_j+i \epsilon}, 
\eqa
where
\bqa 
\label{eq:omega}
\omega_i := q_i/m=(t_i,\vec \rho_i)
\eqa 
and 
\bqa
&& \sigma_1= \omega^2_1-1,~
   \sigma_2= \omega^2_2-1,\nl
&& \sigma_3= \sigma_1+\tau-2 \sqrt{\tau}\, t_1,~
   \sigma_4= \sigma_2+\tau+2 \sqrt{\tau}\, t_2,\nl
&& \sigma_5= (t_1+t_2)^2-\rho^2_1-\rho^2_2-2 \rho_1    \rho_2 c_\theta-\mu_0,  
\eqa
with $\mu_0:= m^2_0/m^2$.
Integrating over the angular variables gives
\bqa
S_2 &=& \frac{4 \pi^2}{m^2}
\int_0^\infty \! d\rho_1 \rho_1
\int_0^\infty \! d\rho_2 \rho_2
\int dt_1 dt_2  \prod_{j=1}^4 \frac{1}{\sigma_j+i \epsilon} \nl
&&\times
\ln   \frac{(t_1+t_2)^2-(\rho_1-\rho_2)^2-\mu_0
+i\epsilon}{(t_1+t_2)^2-(\rho_1+\rho_2)^2-\mu_0+i\epsilon}.
\eqa
The integration over $t_1$ and $t_2$ is trivial and produces
\bqa
\label{eq:S}
S_2 &=& \frac{2 \pi^4}{m^2 \tau} \!\sum_{\lambda_{1,2}= \pm}\!
\int^\infty_{-\infty} \!\!dr_1 
\int^\infty_{-\infty} \!\!dr_2\,
\frac{F(r_1,r_2,\lambda_1,\lambda_2)}{(r_1-i \epsilon) (r_2-i \epsilon)}, 
\eqa
where
\bqa
&& F(r_1,r_2,\lambda_1,\lambda_2)= \lambda_1 \lambda_2\,
\theta(A_1-1) \theta(A_2-1) \nl
&&~\times \ln 
        \frac{r_1+r_2+\sqrt{(\sqrt{A^2_1-1}+\sqrt{A^2_2-1})^2+\mu_0}
 -i \epsilon}{r_1+r_2+\sqrt{(\sqrt{A^2_1-1}-\sqrt{A^2_2-1})^2+\mu_0}-i \epsilon} \nonumber
\eqa
and $A_i := r_i+\lambda_i \sqrt{\tau}/2$.
Threshold singularities at $r_{1,2}= i \epsilon$ are present when $\lambda_{1,2}= +1$ if $\sqrt{\tau} > 2$. When $m_0= 0$,
a two-dimensional implementation of the method of Sect.~\ref{sec:2} gives the results reported in Table \ref{tab:5d}.
Larger MC errors correspond to smaller values of $\rho$. However, we observe that when $\mu_0 \ne 0$ this effect is mitigated. For instance, a $10^{9}$ MC-point estimate with $\rho= .1$ gives
\bqa
-\frac{m^2\tau}{\pi^4} S_2(m,m,m)= 8.582(6) -i\, 2.706(4).
\eqa

\begin{table}
\caption{
The two-loop integral \eqref{eq:S} with $m_0=0$ multiplied
by  $-m^2\tau/\pi^4$ for several values of $\rho := 4/\tau $. Numbers obtained with
$10^{9}$ ($10^{10}$) MC points when $\rho > 1$ ($\rho < 1$). The analytic result is taken from \cite{Broadhurst:1987ei}. MC errors between parentheses.}
\label{tab:5d}       
\begin{tabular}{rll}  
\hline\noalign{\smallskip}
 $\rho$   & MC result & Analytic result \\
\noalign{\smallskip}\hline\noalign{\smallskip}
    .1  & 8.49(1)     $-$$i$ 1.94(2)   & 8.495   $-$$i$ 1.927\\ 
    .3  & 9.34(1)     $-$$i$ 5.47(2)   & 9.340   $-$$i$ 5.460\\  
    .5  & 9.19(1)     $-$$i$ 9.71(1)   & 9.195   $-$$i$ 9.716\\  
    .7  & 7.39(1)     $-$$i$ 15.79(1)  & 7.396   $-$$i$ 15.783 \\
    .9  & -1.03(2)    $-$$i$ 27.591(8) & -1.061  $-$$i$ 27.581 \\  
    1.1 & -15.538(2)  $-$$i$ 1.8314(4)$\times 10^{-5}$   & -15.540 +$i$ 0\\  
    1.3 &  -7.9915(8) $-$$i$ 5.1218(7)$\times 10^{-6}$   & -7.9921  +$i$ 0\\  
    1.5 &  -5.5608(6) $-$$i$ 2.9000(4)$\times 10^{-6}$   &  -5.5614 +$i$ 0\\
    1.7 &  -4.2990(5) $-$$i$ 2.0139(3)$\times 10^{-6}$   &  -4.2996 +$i$ 0\\
    1.9 &  -3.5153(5) $-$$i$ 1.5412(2)$\times 10^{-6}$   &  -3.5157 +$i$ 0\\
\noalign{\smallskip}\hline
\end{tabular}
\end{table}

\section{Gluing together lower-loop structures}
\label{sec:4}
Here we show how higher-loop integrals can be expressed in terms of lower-loop building blocks. Throughout this section dimensionful quantities are rescaled by an arbitrary mass $m$, so that loop momenta are written as in \eqref{eq:omega} and, in particular, 
\bqa
\omega := q/m= (t,\rho c_\theta,\rho s_\theta
s_\phi,\rho s_\theta c_\phi).
\eqa
Furthermore, we define
\bqa
\begin{tabular}{lll}
 $\mu_i := m_i^2/m^2$, & \hskip -5pt 
$\tau := P^2/m^2$, & \hskip -5pt
$\chi:= (p_2-p_3)^2/m^2$, \\
$\tau_i := p^2_i/m^2$, &\hskip -5pt 
$\tau_{ij} := \tau_i-\tau_j$, & \hskip -5pt 
$\lambda_{ij} := \lambda(\tau,\tau_i,\tau_j)$, \\
$k^2 := \lambda_{12}/\tau^2$, &\hskip -5pt 
$k_{\pm} := \sqrt{k^2\pm i \epsilon}$, &\hskip -5pt 
$(k^\prime)^2 :=  \lambda_{34}/\tau^2$,  \\
\end{tabular}
\nonumber
\eqa
and study cases up to a $P := p_1+p_2 \to p_3+p_4$ kinematics of the form
\bqa
\label{eq:kin}
p_1 &=& \frac{m}{2\sqrt{\tau}}
(\tau+\tau_{12},\lambda_{12}^{\frac{1}{2}},0,0), \nl 
p_2 &=& \frac{m}{2\sqrt{\tau}}
(\tau-\tau_{12},-\lambda_{12}^{\frac{1}{2}},0,0),
\nl
p_3 &=& \frac{m}{2 \sqrt{\tau}}
(\tau+\tau_{34},
\lambda_{34}^{\frac{1}{2}}\cos\theta_{13},
\lambda_{34}^{\frac{1}{2}}\sin\theta_{13},0), 
\nl 
p_4 &=& \frac{m}{2 \sqrt{\tau}}
(\tau-\tau_{34},
-\lambda_{34}^{\frac{1}{2}}\cos\theta_{13},
-\lambda_{34}^{\frac{1}{2}}\sin\theta_{13},0).
\eqa
Rescaled propagators belonging to the loop momentum $q$ are denoted by
\bqa
\begin{tabular}{ll}
  $\displaystyle \sigma_0:= {q^2}/{m^2}-\mu_0$,   & $\displaystyle \sigma_1:= {(q-P)^2}/{m^2}-\mu_1$, \\
 $\displaystyle \sigma_2 := {(q-p_2)^2}/{m^2}-\mu_2$, &
 $\displaystyle \sigma_3 := {(q-p_3)^2}/{m^2}-\mu_3$. 
\end{tabular} \nonumber
\eqa
In addition, we define
\bqa
\label{eq:lambda}
\lambda:= 
\lambda(\tau,\sigma_0+\mu_0,\sigma_1+\mu_1),
\eqa
and \footnote{\, $\sigma^{ij}_{a,b,c,d}$ are the invariants
$(q-p_{1,2,3,4})^2/m^2$ computed at values of $t$ and $\rho$ satisfying the conditions 
$
t^2-\rho^2= \sigma_i+\mu_i$ and
$(t-\sqrt{\tau})^2-\rho^2= \sigma_j+\mu_j$.
}
\bqa
\label{eq:sabcd}
\sigma^{ij}_a&:=& \tau_1-\tau+\sigma_j+\mu_j 
+\frac{k c_\theta}{2}\lambda^{\frac{1}{2}}
(\tau,\sigma_i+ \mu_i,\sigma_j+\mu_j)
\nl
&& +\frac{1}{2}
\big(1- {\tau_{12}}/{\tau}\big)
(\tau+\sigma_i+\mu_i-\sigma_j-\mu_j), \nl
\sigma^{ij}_b&:=&
\sigma_i+\mu_i+\sigma_j+\mu_j+\tau_1+\tau_2-\tau-\sigma^{ij}_a, \nl
\sigma^{ij}_c &:=& \tau_3+\sigma_i+\mu_i 
+\frac{k^\prime}{2}\lambda^{\frac{1}{2}}
(\tau,\sigma_i+ \mu_i,\sigma_j+\mu_j) \nl
&& \times \left(c_\theta \cos \theta_{13}+ s_\theta s_\phi \sin \theta_{13} 
\right) \nl
&& -\frac{1}{2}
\big(1+ {\tau_{34}}/{\tau}\big)
(\tau+\sigma_i+\mu_i-\sigma_j-\mu_j), \nl
\sigma^{ij}_d&:=&
\sigma_i+\mu_i+\sigma_j+\mu_j+\tau_3+\tau_4-\tau-\sigma^{ij}_c.
\eqa

The essence of the procedure is to use $\sigma_0$ and $\sigma_1$ as integration variables of the method of Sect.~\ref{sec:2}.
This is achieved by multiplying the integrand by
\bqa 
\label{eq:one}
1 = \dashint d \sigma_0 \dashint  d\sigma_1\, \Delta(\sigma_0,\sigma_1,\rho,t),
\eqa 
where
\bqa
\label{eq:Delta}
&& \Delta(\sigma_0,\sigma_1,\rho,t) :=\delta(\sigma_0+\mu_0+\rho^2-t^2) \nl 
&&~~~~~~~~\times \delta(\sigma_1-\sigma_0 +\mu_1-\mu_0-\tau+2 \sqrt{\tau} t).
\eqa
This gives rise to the appearance of the following three functionals,
\bqa
\label{eq:phis}
\Phi^{(\mu_0,\mu_1)}_1[J_1] &:=& \int d^4 \omega\,J_1\,
\Delta(\sigma_0,\sigma_1,\rho,t), \nl
\Phi^{(\mu_0,\mu_1,\mu_j)}_j[J_j] &:=& \int d^4 \omega\,
\frac{J_j}{\sigma_j+ i \epsilon}\,
\Delta(\sigma_0,\sigma_1,\rho,t),
\eqa
where $j=2,3$.
Assuming $J_1$ independent of any angular variable and $J_2$ ($J_3$) independent of $\theta$ ($\phi$) allows one to compute the functionals  once for all. $\Phi_1$ reads
\bqa
\label{eq:phi1}
\Phi^{(\mu_0,\mu_1)}_1[J_1]= \frac{\pi}{2 \tau}
\lambda^{\frac{1}{2}} \theta(\lambda) J_1.
\eqa
As for $\Phi_2$, one has 
\bqa 
\label{eq:phi2}
\Phi^{(\mu_0,\mu_1\mu_2)}_2[J_2]&=& \frac{1}{4 \tau } \theta(\lambda)
\Bigg[
\frac{1}{k_+} \ln \left(1+\frac{k_+ \lambda^{\frac{1}{2}}}{A_2+i \epsilon}\right) \\
&&-
\frac{1}{k_-} \ln \left(1-\frac{k_- \lambda^{\frac{1}{2}}}{A_2+i \epsilon}\right) 
\Bigg] \int_0^{2 \pi} d \phi\, J_2. \nonumber
\eqa 
with
\bqa
A_2 &:=& (\sigma_0+\mu_0)
\bigg(1+\frac{\tau_{12}}{\tau}\bigg)
    + (\sigma_1+\mu_1)\bigg(1-\frac{\tau_{12}}{\tau}\bigg)  \nl
     &&+\tau_1+\tau_2-\tau -2\mu_2.
\eqa
Note that \eqref{eq:phi1} and \eqref{eq:phi2} have 
been analytically continued to configurations
with any sign of $\tau$, $\tau_1$,  $\tau_2$, $\lambda_{12}$. 
Finally
\bqa 
\label{eq:phi3}
\Phi^{(\mu_0,\mu_1,\mu_3)}_3[J_3]= \frac{\pi}{2 \tau} \lambda^{\frac{1}{2}} \theta(\lambda) 
\int_{-1}^1{ dc_\theta \frac{J_3}{A_3} \frac{1}{\sqrt{1-\frac{B_3}{A^2_3}}}},
\eqa  
in which
\bqa
A_3 &:=&  (\sigma_0+\mu_0)
\bigg(1-\frac{\tau_{34}}{\tau}\bigg)
    + (\sigma_1+\mu_1)\bigg(1+\frac{\tau_{34}}{\tau}\bigg)  \nl
     &&+\tau_3+\tau_4-\tau -2\mu_3
     + \lambda^{\frac{1}{2}}k^\prime c_\theta \cos \theta_{13}+i \epsilon,
\nl
B_3 &:=& \frac{\lambda \lambda_{34}}{\tau^2} s^2_\theta
\sin^2 \theta_{13}.
\eqa 

\begin{figure}
\vskip -4.6cm
\hskip -3.9cm
\includegraphics[width=6.5in]{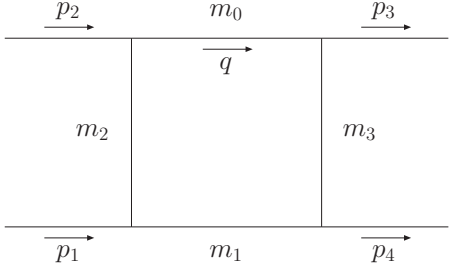}
\vskip -4.2cm
\caption{The scalar four-point one-loop function 
$D(p^2_1,p^2_2,p^2_3,p^2_4,(p_1\!\!+\!\!p_2)^2,(p_2\!\!-\!\!p_3)^2,m^2_1,m^2_2,m^2_0,m^2_3)$
with arbitrary kinematics and  masses.}
\label{fig:box}   
\end{figure}

\subsection{One-loop examples}
\label{sec:ole}
To elucidate the procedure, we first consider gluing tree-level structures to compute the three-point function of Fig.~\ref{fig:1}\,--\,with arbitrary kinematics and masses\,--\,and the box diagram of Fig.~\ref{fig:box} with 
$p^2_i= 0$ and $m_0=m_1=m_2=m_3=m$, which we dub $C$ and $D_0$, respectively. Equations \eqref{eq:one}, \eqref{eq:phi2} and
\eqref{eq:phi3} produce
\bqa
\label{eq:C0}
m^2 C 
&=& \dashint \prod_{j=0}^{1}  \left(
\frac{d \sigma_j}{\sigma_j+i \epsilon} \right)\Phi^{(\mu_0,\mu_1)}_2[1], \\
\label{eq:D}
m^4 D_0 &=& \dashint \prod_{j=0}^{1}  \left(
\frac{d \sigma_j}{\sigma_j+i \epsilon} \right)\Phi^{(\mu_0,\mu_1,\mu_3)}_3\left[\frac{1}{\sigma_2+i \epsilon}\right]. 
\eqa
In \eqref{eq:C0} the integration over the azimuth angle $\phi$ is trivial,
\bqa
\label{eq:phi2c}
\Phi^{(\mu_0,\mu_1)}_2[1]= \frac{\pi}{2 \tau} \theta(\lambda)
\sum_{j= \pm}
\left[
\frac{1}{jk_j} \ln \left(1+\frac{jk_j \lambda^{\frac{1}{2}}}{A_2+ i \epsilon} \right)
\right],
\eqa
while the integral over $c_\theta$ in \eqref{eq:D} has to be dealt with numerically using the method of Sect.~\ref{sec:2}, which gives 
\bqa
\label{eq:phi3d}
\Phi^{(\mu_0,\mu_1,\mu_3)}_3\left[\frac{1}{\sigma_2+i \epsilon}\right] &=&
\frac{\pi}{\tau} \theta(\lambda_0)
\dashint \frac{d \sigma_2}{\sigma_2+ i \epsilon} \nl
&& \times \frac{\theta(\sigma_2- \sigma^-_2) 
\theta(\sigma^+_2- \sigma_2)}{A_0 \sqrt{1-\frac{B_0}{A^2_0}}}, 
\eqa
where
\bqa
\label{eq:lamdda0}
\sigma^\pm_2 &=& \frac{\sigma_\tau\pm\lambda_0^{\frac{1}{2}}}{2},~\sigma_\tau:= \sigma_0+\sigma_1-\tau, \nl
A_0 &=& 2 \left[ \sigma_2-
\frac{\chi}{\tau}\left(\sigma_\tau-2 \sigma_2 \right)
\right]+i \epsilon, \nl
B_0 &=& -16\frac{\chi}{\tau}
\left(1+\frac{\chi}{\tau}\right)
\left[\sigma_2(\sigma_\tau-\sigma_2)-\tau-\sigma_0\sigma_1 \right],
\nl
\lambda_0 &=&  \lambda(\tau,\sigma_0+1,\sigma_1+1).
\eqa
\paragraph{Improving the numerical accuracy} 
$~$ \vskip 5pt
\noindent When inserted in \eqref{eq:C0} and \eqref{eq:D}, equations \eqref{eq:phi2c} and \eqref{eq:phi3d} could potentially produce inaccurate results 
when strong cancellations are expected among different integration regions. This  happens if  
\bqa
\label{eq:ab}
\mbox{(a)} && \mbox{the integrands do not vanish fast enough at large} \nl
&& \mbox{values of $\sigma_{0,1}$}; \nl
\mbox{(b)} && \mbox{$\tau$ is small.}
\eqa 
Note that case (a) is relevant to $C$ but not to $D_0$, since 
\bqa
\label{eq:phi2lim}
&&\lim_{\sigma_{0,1} \to \infty}  \Phi^{(\mu_0,\mu_1)}_2[1]
\sim{\rm constant}\,, \\
&&\lim_{\sigma_{0,1} \to \infty}  
\Phi^{(\mu_0,\mu_1,\mu_3)}_3\left[\frac{1}{\sigma_2+i \epsilon}\right]
\sim \frac{1}{\sigma_{0,1}},
\eqa
while (b) applies to both $C$ and $D_0$ due to the common $1/\tau$ prefactor. 
In the following paragraphs we illustrate how numerical inaccuracies caused by the configurations (a) and (b) in \eqref{eq:ab} can be circumvented.

As for case (a), a preliminary analysis is in order to understand the mechanism that makes $C$ finite despite \eqref{eq:phi2lim}. \footnote{The principal value integrations in \eqref{eq:C0} are not sufficient to regularize the large $\sigma_{0,1}$ behaviour. In fact,
$\Phi^{(\mu_0,\mu_1)}_2[1]$ approaches
two different constants when  $\sigma_{0,1} \to \infty$ or $\sigma_{0,1} \to -\infty$, so that no cancellation is possible.} We define
\bqa
\label{eq:s01}
\sigma_{10} := \sigma_1-\sigma_0,
\eqa
in terms of which
\bqa
m^2 C = \lim_{\Lambda \to \infty} 
\int_{-\Lambda}^{\Lambda} d \sigma_{10}
\int d \sigma_0 \frac{\Phi^{(\mu_0,\mu_1)}_2[1]}{(\sigma_0+i \epsilon)(\sigma_0+\sigma_{10}+ i \epsilon)}. \nonumber
\eqa
Now the $\sigma_0$ integral is convergent by power counting and $\lambda$ in $\eqref{eq:lambda}$ behaves as 
\bqa
\lim_{\sigma_{10} \to \infty} \lambda \sim \tilde \lambda:= (\sigma_{10}+\beta)^2-\alpha,
\eqa
where $\alpha$ and $\beta$ are constants.
Replacing $\lambda$ with $\tilde \lambda$ in \eqref{eq:phi2c} produces an integrand in which all branch points and poles are located in the lower $\sigma_0$ complex half-plane. As a result, the integral over $\sigma_0$ approaches zero when $\sigma_{10} \to \pm \infty$, so that the $\Lambda \to \infty$ limit exists.
This same reasoning allows one to construct a class of vanishing integrals defined as
\bqa
m^2 \tilde C(\alpha,\beta,\Lambda_0) :=
\int \frac{d^4 \omega\, \,\theta(|\sigma_{10}|-\Lambda_0)}{(\sigma_0+i \epsilon)(\tilde \sigma_1+i \epsilon) (\tilde \sigma_2+i \epsilon)},
\eqa
with
\bqa
\label{eq:ss}
m^2\tilde \sigma_1 &:=& q^2+P^2  -m^2_1 -2 (\tilde q \cdot P), \nl
m^2\tilde \sigma_2 &:=& q^2+p^2_2-m^2_2 -2 (\tilde q \cdot p_2),
\eqa
where $\tilde q$ is the asymptotic $|\sigma_{10}| \to \infty$ limit of $q$,
\bqa
\frac{\tilde q}{m} := ( \tilde t, \vec{\tilde \rho}),\,
\tilde t:= -\frac{\sigma_{10}+\beta}{2 \sqrt{\tau}},\,
\tilde \rho := \frac{{\tilde \lambda}^{\frac{1}{2}}}{2 \sqrt{\tau}}.
\eqa
Again, all cuts and poles lie in the lower $\sigma_0$ complex half-plane, so that 
\bqa
\label{eq:C0tilde}
\tilde C(\alpha,\beta,\Lambda_0)= 0.
\eqa
Now $\alpha$ and $\beta$ can be set to obtain a local cancellation of the problematic large $\sigma_{10}$ configurations. \footnote{Additionally, $\Lambda_0$ can be used to control when such a local subtraction has to be performed.}
An explicit calculation with $\tau_1= \tau_2= 0$ gives
\bqa
\label{eq:albet}
\alpha= 4 \tau \mu_2,~\beta= \mu_1-\mu_0.
\eqa
When $\tau_i \ne 0$ the accuracy of \eqref{eq:C0} is improved by the non-vanishing external masses, so that \eqref{eq:albet} is relevant to this case as well.
In summary,  the formula
\bqa
\label{eq:Csub}
m^2C= m^2C- m^2\tilde C(\alpha,\beta,\Lambda_0)
\eqa
produces numerically  stable  results with $\alpha$ and $\beta$ given in \eqref{eq:albet} when the same sequence of $\sigma_0$ and $\sigma_1$ values are used in both $C$ and $\tilde C$.

It turns out that the configurations of type (b) of $C$ are also cured by the subtraction in \eqref{eq:Csub}. Thus, we are only left with the discussion of the case (b) for $D_0$.
In the  $\tau \to 0$ region it is convenient to give up the exact formula and use, instead, a few terms of a Taylor expansion in $\tau$ and $\chi$
obtained with the method given in \ref{app:a},
\bqa
\label{eq:tayD0}
\frac{D_0}{i \pi^2} &=&
\frac{1}{6 m^4}\left[
 1
+\frac{\tau+\chi}{10}
+\frac{\tau^2+\chi^2}{70}
+\frac{\tau \chi}{140} \right.\nl
&&\left.+\frac{\tau^3+\chi^3}{420}
+\frac{\tau \chi (\tau+\chi)}{1260} + {\cal O} \left(\tau^4\right) \right]. 
\eqa
By doing that, it is easy to find a value of $\tau$ below which the exact result is well approximated by \eqref{eq:tayD0}, and above which \eqref{eq:D} is accurate.
\paragraph{Results} 
$~$ \vskip 5pt
\noindent Here we present the numerical
outcome of a MC based on \eqref{eq:C0}, \eqref{eq:D}, \eqref{eq:Csub} and \eqref{eq:tayD0}. 

\begin{figure}
\vskip -4.3cm
\hskip -3.1cm
\includegraphics[width=5.7in]{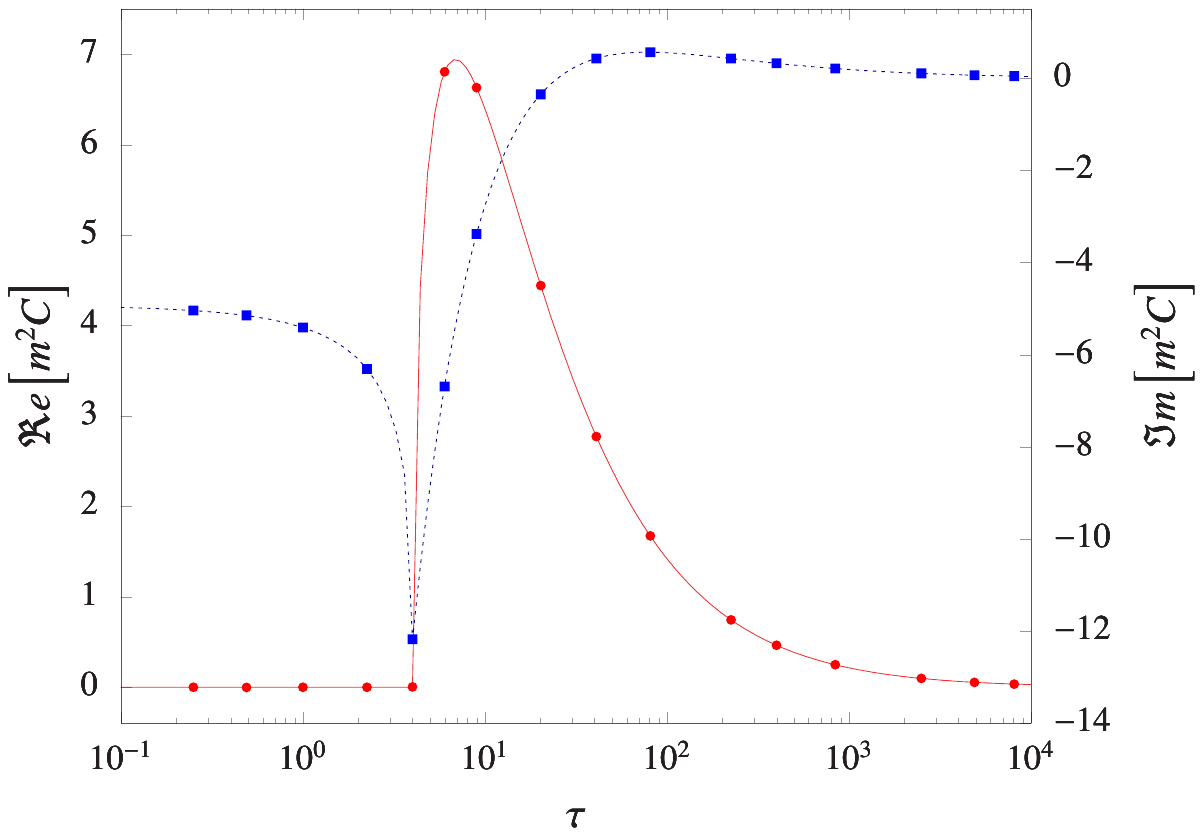}
\vskip -10.cm
\caption{MC estimates of the real (red bullets) and imaginary (blue squares) part of \eqref{eq:Csub}, evaluated with $m_i= m$, $p^2_i= 0$ and $\Lambda_0= 0$, compared to the analytic output of \cite{vanHameren:2010cp}. 
Results obtained with $10^{9}$ MC shots per point.}
\label{fig:2}  
\end{figure}

\begin{figure}
\vskip -4.3cm
\hskip -3.1cm
\includegraphics[width=5.7in]{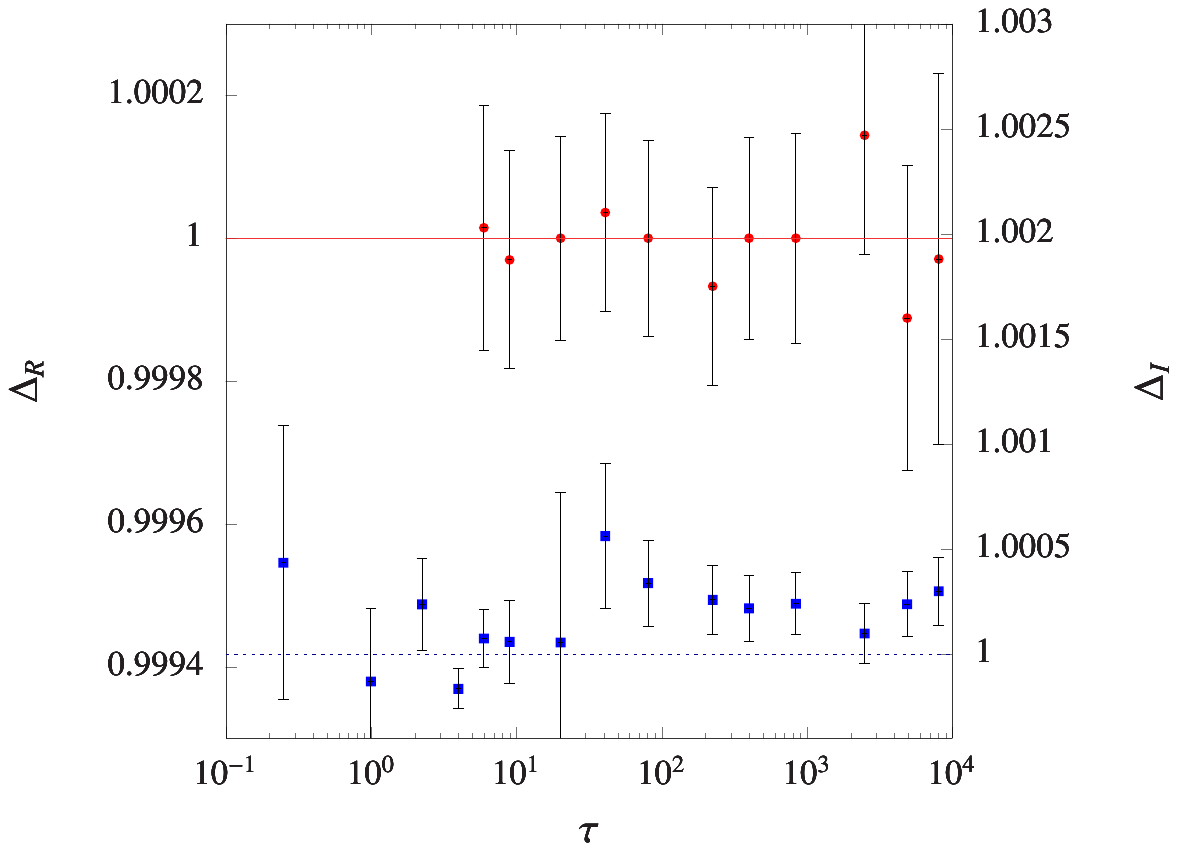}
\vskip -10cm
\caption{The relative difference between the MC and the non-zero analytic results of Fig.~\ref{fig:2} as defined in \eqref{eq:Delta_RI}.
Red bullets (blue squares) refer to $\Delta_R$ ($\Delta_I$).}
\label{fig:3}   
\end{figure}

The
results for $C(P^2,0,0,m,m,m)$ are shown in Figs. \ref{fig:2} and \ref{fig:3}. In the latter, the relative difference between the MC and the analytic (AN) non-zero results of the former is plotted in terms of
\bqa
\label{eq:Delta_RI}
\Delta_{R,I} := 1+\left(\frac{{\rm MC}-{\rm AN}}{\rm AN}
\right)_{R,I},
\eqa
where $R$ and $I$ refer to the real and imaginary parts, respectively.    
With the given MC statistics, the analytic result is reproduced by 
\eqref{eq:Csub} within a few parts in $10^4$,\footnote{Similar results are obtained when $\tau < 0$. For instance, with $\tau= -10^3$ Eq.~\eqref{eq:Csub} gives  $m^2 C= 1(1)\times 10^{-6} -i\,0.23566(5)$, to be compared to the analytic value $-i\, 0.23561$.} which is an accuracy comparable to the one of Table \ref{tab:5c}, but obtained with 50 times more points. The main reason for this difference is that the one-dimensional representation \eqref{eq:c0III} is now replaced by the twofold integration \eqref{eq:C0}. On the other hand, the gluing algorithm is highly modular and can be extended to more complex situations, as we will see in the next subsection.

Table \ref{tab:3} displays our results for $D_0$ as a function of $\tau$ and the scattering angle
$
-{\chi}/{\tau}= \left(1+\cos \theta_{13} \right)/2
$.
We use the MC evaluation of \eqref{eq:D} above $\tau= 1$ and
the Taylor expansion of \eqref{eq:tayD0} when
$\tau < 1$. In the former case, an accuracy of a few parts 
in $10^3$ is reached with $10^9$ MC points.

\begin{table*}
\caption{Numerical estimates of $m^4 D_0$ compared to the analytic result of \cite{vanHameren:2010cp}. Equation \eqref{eq:D} [\eqref{eq:tayD0}] is used if
$\tau > 1$ [$\tau < 1$].  When $\tau > 1$, $10^9$ MC points are used. Statistical errors between parentheses.}
\label{tab:3}       
\begin{tabular}{rrll}
\hline\noalign{\smallskip}
 $\tau$  & $-\chi/\tau$ & Numerical result & Analytic result \\
\noalign{\smallskip}\hline\noalign{\smallskip}
    0.8 &  0.1   &  1.7798  &  1.7801\\ 
    0.8 &  0.5   &  1.7274  &  1.7277\\ 
    0.8 &  0.9   &  1.6790  &  1.6795\\    
    10 &   0.1   & -2.185(9) +$i$ 2.64(9)$\times 10^{-1}$   & -2.1871  +$i$ 2.7316$\times 10^{-1}$ \\ 
    10 &   0.5   &  -1.642(4) +$i$   4.4(4)$\times 10^{-2}$  & -1.6423  +$i$   4.2586$\times 10^{-2}$ \\ 
    10 &   0.9   & -1.339(4) $-$$i$  3.6(4)$\times 10^{-2}$  & -1.3368   $-$$i$  4.0162$\times 10^{-2}$ \\ 
    100 &   0.1   & -1.297(3)$\times 10^{-1}$ $-$$i$ 1.253(2)$\times 10^{-1}$   & -1.2993$\times 10^{-1}$  $-$$i$ 1.2556$\times 10^{-1}$ \\ 
    100 &   0.5   &  -4.761(8)$\times 10^{-2}$ $-$$i$ 5.176(9)$\times 10^{-2}$  &  -4.7572$\times 10^{-2}$ $-$$i$  5.1670$\times 10^{-2}$ \\ 
    100 &   0.9   & -3.090(8)$\times 10^{-2}$ $-$$i$ 3.447(8)$\times 10^{-2}$   & -3.0812$\times 10^{-2}$  $-$$i$ 3.4615$\times 10^{-2}$ \\  
    1000 &   0.1   &  -2.803(5)$\times 10^{-3}$ $-$$i$ 5.277(5)$\times 10^{-3}$  &  -2.8153$\times 10^{-3}$ $-$$i$ 5.2737$\times 10^{-3}$ \\ 
    1000 &   0.5   & -7.66(2)$\times 10^{-4}$ $-$$i$  1.501(2)$\times 10^{-3}$  & -7.6923$\times 10^{-4}$  $-$$i$ 1.4986$\times 10^{-3}$ \\ 
    1000 &   0.9   &  -4.68(2)$\times 10^{-4}$ $-$$i$ 9.20(2)$\times 10^{-4}$  &  -4.6848$\times 10^{-4}$ $-$$i$ 9.2226$\times 10^{-4}$ \\     
\noalign{\smallskip}\hline
\end{tabular}
\end{table*}

\subsection{Two and three-loop examples}
\paragraph{Two-loop self-energy}
$~$ \vskip 5pt
\noindent The two-loop diagram of Fig.~\ref{fig:1i} can be easily obtained by gluing together a one-loop triangle and a tree-level structure. For instance, with $m_0=m_1=0$ one has
\bqa
\label{eq:Sm00}
m^2 S_2(m,0,0)
&=& \dashint \prod_{j=0}^{1}  \left(
\frac{d \sigma_j}{\sigma_j+i \epsilon} \right) \\
&&\!\times \Phi^{(1,1)}_1[C(\tau,\sigma_1+1,\sigma_0+1,0,0,0)]. \nonumber
\eqa
Eq.~\eqref{eq:Sm00} suffers the inaccuracies of type (a) of
\eqref{eq:ab}.
To cure this, we use the same strategy described in
Sect.~\ref{sec:ole}. First, we consider an explicit representation of the triangle as a function of $\sigma_0$ and $\sigma_{10}$ in
\eqref{eq:s01},
\bqa
\label{eq:Cint}
C(\tau,\sigma_1+1,\sigma_0+1,0,0,0) = 2 \frac{i \pi^2}{\lambda_0^{\frac{1}{2}}} F(\sigma_0,\sigma_{10}),
\eqa
with $\lambda_0$ in \eqref{eq:lamdda0},
\bqa
\label{eq:F2}
&&F(\sigma_0,\sigma_{10})= \nl
&&~~~\sum_{j=1,2} (-1)^{j} \Bigg\{
 {\rm Li}_2 \Big(\frac{2 \tau}{v_j}\Big)
+ \ln\Big(\frac{2 \tau}{v_j}\Big) 
  \ln\Big(1-\frac{2\tau}{v_j}\Big) \nl
&&~~~-\frac{1}{4} \ln^2 \Big(\frac{2\tau}{v_j}-1\Big)  
-\frac{1}{4}  \ln^2 \Big(\frac{u_j}{v_j} \Big) \Bigg\}, 
\eqa
and
\bqa
\begin{tabular}{ll}
  $\!\!u_1= \tau-\sigma_{10}+\lambda_0^{\frac{1}{2}}-i \epsilon \tau$,
 &$\!u_2= \tau-\sigma_{10}-\lambda_0^{\frac{1}{2}}+i \epsilon \tau$ \\ 
  $\!\!v_1= \tau+\sigma_{10}+\lambda_0^{\frac{1}{2}}-i \epsilon \tau$,
 &$\!v_2= \tau+\sigma_{10}-\lambda_0^{\frac{1}{2}}+i \epsilon \tau$. 
\end{tabular}
\nonumber
\eqa
From this, it is easy to determine a  $|\sigma_{10}| \to \infty$ asymptotic approximant 
of   \eqref{eq:Cint}  that gives zero upon integration over $\sigma_{0,1}$. The results reads
\footnote{As in the case of \eqref{eq:C0tilde}, \eqref{eq:Stilde} is proven by observing that
all  cuts  and  poles  lie  in  the  lower $\sigma_0 $ complex half-plane.}
\bqa
\label{eq:Stilde}
m^2 \tilde S_2(\Lambda_c) &:=& \frac{i \pi^3}{\tau}
 \dashint \prod_{j=0}^{1}  \left(
\frac{d \sigma_j}{\sigma_j+i \epsilon} \right) \theta(\tilde \lambda_0)  \nl
&& \times \theta\Big(\Lambda_c^2-\frac{\tau^2}{\tau^2+
\sigma^2_{10}}\Big) \tilde F(\sigma_0,\sigma_{10})= 0,
\eqa
where $\tilde \lambda_0:= (\sigma_{10}-\tau)^2-4 \tau$, $\Lambda_c$ is arbitrary and $\tilde F(\sigma_0,\sigma_{10})$ is constructed by replacing in \eqref{eq:F2} the $u_j, v_j$  with their asymptotic counterparts  $\tilde u_j, \tilde v_j$  defined as
\bqa
\tilde u_1 &=& \tilde a\,\theta( \sigma_{10})
+\tilde b\,\theta(-\sigma_{10})-i \epsilon \tau, \nl
\tilde u_2 &=& 
 \tilde b\,\theta( \sigma_{10})
 +\tilde a \,\theta(-\sigma_{10})+i \epsilon \tau, \nl
\tilde v_1 &=& 
2 \sigma_{10}\,\theta( \sigma_{10})
+ \tilde c
\,\theta(-\sigma_{10})-i \epsilon \tau, \nl
\tilde v_2 &=& \tilde c
\,\theta( \sigma_{10})
+ 2 \sigma_{10}\,\theta(-\sigma_{10})+i \epsilon \tau, 
\eqa
with
\bqa
\tilde a &:=& -\frac{2 \tau}{\sigma_{10}}(\sigma_0+1),~
\tilde b := 2(\tau-\sigma_{10}), \nl
\tilde c &:=&  \frac{2 \tau}{\sigma_{10}}(\sigma_{10}+\sigma_0+1).
\eqa
Finally, in the same spirit as \eqref{eq:Csub},  we rewrite
\bqa
\label{eq:Ssub}
S_2(m,0,0)= S_2(m,0,0)-\tilde S_2(\Lambda_c),
\eqa
where we understand a local subtraction of the large $|\sigma_{10}|$ configurations.

\begin{figure}
\vskip -4.8cm
\hskip -5.3cm
\includegraphics[width=6.5in]{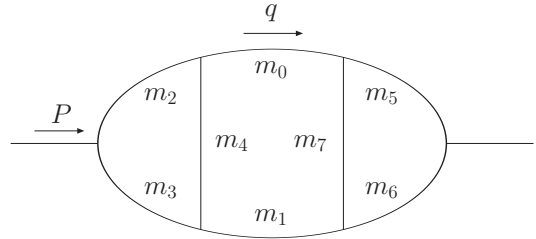}
\vskip -4.8cm
\caption{The three-loop self-energy diagram $S_3(\{\mu_k\})$ with masses $\{m^2_k/m^2\} =\{\mu_k\} :=  \{\mu_0,\mu_1,\mu_2,\mu_3,\mu_4,\mu_5,\mu_6,\mu_7\}
$.}
\label{fig:3loop}   
\end{figure}

\begin{figure}
\vskip -4.3cm
\hskip -3.2cm
\includegraphics[width=5.7in]{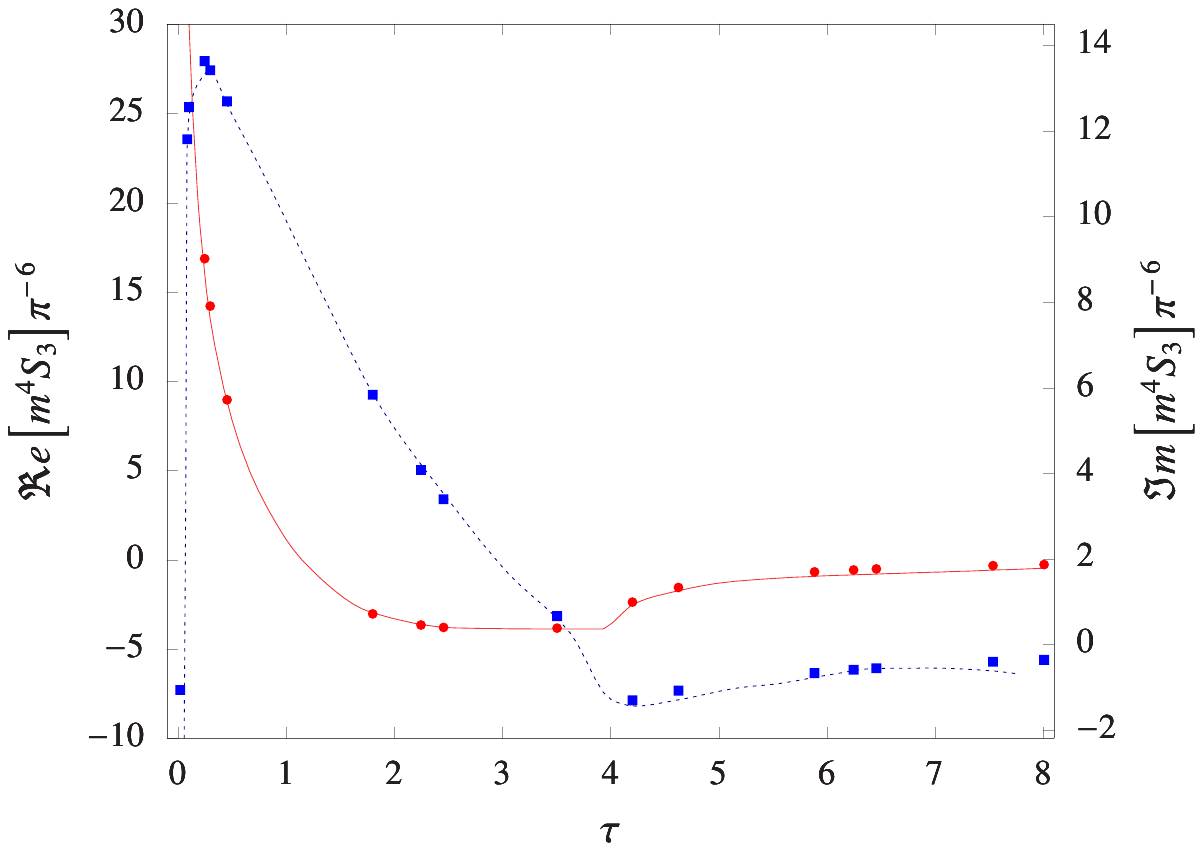}
\vskip -10.1cm
\caption{The three-loop integral $m^4 S_3(1,1,0,0,0,0,0,0)$ as a function of $\tau$. 
Red bullets (blue squares) refer to the real (imaginary) part computed with $10^{9}$ MC shots per point.
The solid and dashed lines are obtained from \cite{Ghinculov:1996vd}. 
To compare, we  divided (not multiplied, as stated in \cite{Ghinculov:1996vd}) our equation \eqref{eq:T} by $\pi^6$.
}
\label{fig:Tplot}   
\end{figure}

In Table \ref{tab:S} we present our numerical estimates based on
\eqref{eq:Ssub} with $\Lambda_c= 1/2$.
An accuracy of the order of $10^{-3}$ is achieved with $10^9$ MC points. We also studied the stability of \eqref{eq:Ssub} at small values of $\tau$. For instance, when
$\rho= 100$ ($\tau=0.04$) we obtain
\bqa
\label{eq:S2number}
-m^2\frac{\tau}{\pi^4}
S_2(m,0,0) =-0.2090(2) -i\,0.1263(3).
\eqa
Note that determining $\tilde S_2(\Lambda_c)$ requires an analytic knowledge of the integrand.  When this is not possible, an alternative approach is to cut away the problematic configurations in a controlled manner. For instance, discarding in \eqref{eq:Sm00} integration points with
\bqa
\label{eq:Lambdacut}
\sqrt{\tau^2/(\tau^2+\sigma^2_{10})} < \Lambda
\eqa
is expected to produce an error of ${\cal O}(\Lambda^2)$. Indeed we checked that, with $\Lambda= 0.01$, the fraction of the integral discarded by \eqref{eq:Lambdacut} is always below the errors reported in Table \ref{tab:S}.
\begin{table}
\caption{
The two-loop integral \eqref{eq:Ssub} with $\Lambda_c= 1/2$  multiplied
by  $-m^2\tau/\pi^4$ as a function of $\rho := 4/\tau$. Numbers obtained with $10^{9}$ MC points. Statistical errors between parentheses.}
\label{tab:S}       
\begin{tabular}{rl}
\hline\noalign{\smallskip}
 $\rho$   & MC result \\
\noalign{\smallskip}\hline\noalign{\smallskip}
    -1.5 &   3.325(3)   $-$$i$    1(2)$\times 10^{-3}$   \\
    -.5  &   4.817(4)   +$i$      1(3)$\times 10^{-3}$   \\
    -.1  &   6.297(6)   +$i$      4(7)$\times 10^{-3}$    \\
    -.01 &   7.029(9)   $-$$i$    1(1)$\times 10^{-2}$   \\
    -.001&   7.16(1)    +$i$      1(3)$\times 10^{-2}$   \\
     .001&   7.234(7)   $-$$i$    2(3)$\times 10^{-2}$  \\
     .01 &   7.358(5)   $-$$i$    1.3(2)$\times 10^{-1}$  \\
     .1  &   7.932(3)   $-$$i$    9.30(8)$\times 10^{-1}$  \\
     .5  &   8.990(3)   $-$$i$    3.981(4)    \\
     1.5 &   3.60(2)$\times 10^{-1}$ $-$$i$ 9.464(2)   \\
\noalign{\smallskip}\hline
\end{tabular}
\end{table}
\paragraph{Three-loop self-energy}
$~$ \vskip 5pt
\noindent Our next example is the scalar three-loop self-energy of Fig.~\ref{fig:3loop}, which is attained by gluing together the two triangles
\bqa
C_L&:=& C(\tau,\sigma_1+\mu_1,\sigma_0+\mu_0,\mu_2,\mu_3,\mu_4), \nl
C_R&:=& C(\tau,\sigma_1+\mu_1,\sigma_0+\mu_0,\mu_5,\mu_6,\mu_7),
\eqa
by means of $\Phi_1$,
\bqa
\label{eq:T}
m^4 S_3(\{\mu_k\})
&=& \dashint \prod_{j=0}^{1}  \left(
\frac{d \sigma_j}{\sigma_j+i \epsilon} \right) \Phi^{(\mu_0,\mu_1)}_1[C_L C_R].
\eqa
Now the integrand vanishes fast enough at large $\sigma_{0,1}$, so that no subtraction is needed. Our results for the case 
$\{\mu_k\}= \{1,1,0,0,0,0,0,0\}$ are shown in Fig.~\ref{fig:Tplot}, where we compare
 them with digitized curves obtained from Fig. 4 of \cite{Ghinculov:1996vd}. The agreement is good, but our points tend to overshoot the lines at large $\tau$. However, the quality of the plot in Ref.~\cite{Ghinculov:1996vd} is poor
there and the digitization may be misleading.
Thus, we cross-checked internally our high-energy results by comparing the outcome of two independent MCs based on method 1 and 2 of Sect. \ref{sec:2}, respectively. We did not find any systematic difference
in the range $7 < \tau < 50$. Finally, we observe that dealing with arbitrary mass configurations poses no difficulties whatsoever. For example, when $\{\mu_k\}= \{1,1,2,3,4,5,6,7\}$,
using {\tt OneLOop} \cite{vanHameren:2010cp} to evaluate the $C_{L,R}$ triangles gives, at $\tau= 10$ and with  $10^{9}$ points,
\bqa
\frac{m^4}{\pi^6} S_3(\{\mu_k\})
= 1.1453(8)\times 10^{-1} -i\,4.11(1)\times 10^{-2}. \nonumber
\eqa

\begin{figure}[t]
\vskip -5.3cm
\hskip -5.0cm
\includegraphics[width=6.5in]{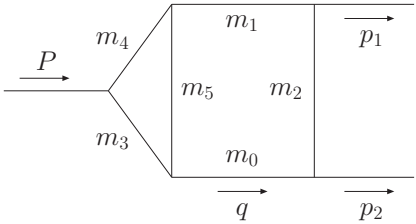}
\vskip -4.5cm
\caption{The planar two-loop vertex $V_2(\{\mu_k\})$ of \eqref{eq:V2} with masses $\{m^2_k/m^2\} =\{\mu_k\} :=  \{\mu_0,\mu_1,\mu_2,\mu_3,\mu_4,\mu_5\}
$.}
\label{fig:V2}   
\end{figure}

\paragraph{Planar two-loop vertex}
$~$ \vskip 5pt
\noindent Consider now the two-loop scalar vertex of Fig.~\ref{fig:V2}. Gluing the one-loop triangle on the left 
\bqa
C_L := C(\tau,\sigma_1+\mu_1,\sigma_0+\mu_0,\mu_3,\mu_4,\mu_5)\nonumber
\eqa
by means of $\Phi_2$ gives
\bqa
\label{eq:V2}
&&m^4 V_2(\{\mu_k\})
= \dashint \prod_{j=0}^{1}  \left(
\frac{d \sigma_j}{\sigma_j+i \epsilon} \right)  \Phi^{(\mu_0,\mu_1,\mu_2)}_2[C_L]. 
\eqa
This representation holds true with any sign of $\tau, \tau_1, \tau_2$ and for any choice of internal masses 
\footnote{{The inclusion of complex masses is in principle possible, although a dedicated study is needed in this case to assess the numerical accuracy, especially for small width-to-mass ratios.}}. In addition, it does not require subtracting large $\sigma_{0,1}$ configurations. In Table \ref{tab:V2} we collect a few results obtained by computing the triangle with {\tt OneLOop}. {The last row refers to the Standard-Model-like $Z\to \nu \bar \nu$ case with $m_0=m_1= m_e$, $m_2=m_3=m_4=M_W$, $m_5= 0$, $P^2= M^2_Z$ and $p^2_{1,2}= 0$. This shows that the MC error is under control also for configurations with large mass gaps.}
Note that $V_2(\{\mu_k\})$ can also be obtained by gluing together the box on the right
$D_R := D(\tau_1,\tau_2,\sigma_3+\mu_3,\sigma_4+\mu_4,\tau,\sigma^{34}_b,\mu_1,\mu_2,\mu_0,\mu_5)$
and the tree-level decay on the left. When 
$k^2$ is greater than zero one has
\bqa 
\label{eq:V2D}
m^4 V_2(\{\mu_k\})
= \dashint \prod_{j=3}^{4}  \left(
\frac{d \sigma_j}{\sigma_j+i \epsilon} \right) \int_{-1}^{1} \frac{d c_\theta}{2} \Phi^{(\mu_3,\mu_4)}_1[D_R]. \nl
\eqa
The loop momentum of the $m_3$ line of Fig.~\ref{fig:V2} flows through just one propagator of $D_R$. Because of that, a ``technical'' cut $\sqrt{\tau^2/(\tau^2+(\sigma_3-\sigma_4)^2)} < \Lambda$ is required to damp the inaccurate large $\sigma_{3,4}$ behaviour of \eqref{eq:V2D}. By power counting, the discarded contribution is of ${\cal O}(\Lambda^4)$. With 
$\Lambda = 0.05$ and $10^9$ MC points  \eqref{eq:V2D} reproduces the numbers of Table \ref{tab:V2} at the percent level. 

\begin{figure}
\vskip -5.05cm
\hskip -5.0cm
\includegraphics[width=6.5in]{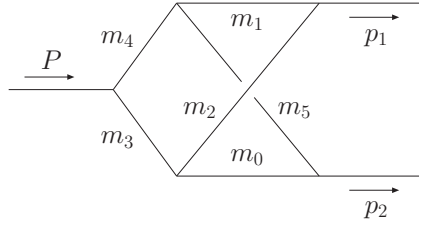}
\vskip -4.4cm
\caption{The non-planar two-loop vertex $V^{np}_2(\{\mu_k\})$ of \eqref{eq:V2np} with masses $\{m^2_k/m^2\} =\{\mu_k\} :=  \{\mu_0,\mu_1,\mu_2,\mu_3,\mu_4,\mu_5\}
$.}
\label{fig:V2np}   
\end{figure}

\begin{table*}
\caption{
The planar two-loop integral \eqref{eq:V2} as a function of arbitrary input parameters. Numbers obtained with $10^{9}$ MC points. Statistical errors in parentheses.}
\label{tab:V2}       
\begin{tabular}{crrcl}
\hline\noalign{\smallskip}
 $\tau$ & $\tau_1$ & $\tau_2$ & $\{\mu_k\}$ & MC result \\
\noalign{\smallskip}\hline\noalign{\smallskip}
 $10$  & $2$ & $3$ & $\{1,2,3,4,5,6\}$  & -2.751(7)                  $-$$i$ 6.729(7) \\
 $-100$& $-4$& $7$ & $\{0,1,12,7,8,9\}$ & -1.025(1)$\times 10^{-1}$ +$i$ 1.5(7)$\times 10^{-4}$\\
 $50$  & $30$& $30$& $\{6,23,2,9,9,9\}$ &  9.307(4)$\times 10^{-1}$  $-$$i$ 1.347(4)$\times 10^{-1}$\\
 $-40$ & $1$ & $1$ & $\{0,0,0,0,0,0\}$  &  5.918(6)                  $-$$i$ 9.51(3)  \\
$1000$ & $0$ & $0$ & $\{1,1,1,1,1,1\}$  & -3.558(4)$\times 10^{-3}$ +$i$ 2.3557(8)$\times 10^{-2}$ \\
{31.9$\times 10^{9}$} & {0} & {0}&
{$\{1,1,24.7\!\times\!10^{9},24.7\!\times\!10^{9},24.7\!\times\!10^{9},0\}$}&
{-2.5(8)$\times 10^{-22}$ $-$$i$ 1.7947(1)$\times 10^{-19}$}\\
\noalign{\smallskip}\hline
\end{tabular}
\end{table*}
\paragraph{Non-planar two-loop vertex}
$~$ \vskip 5pt
\noindent The same reasoning leading to \eqref{eq:V2D} allows one to compute the non-planar two-loop vertex of Fig.~\ref{fig:V2np},
\bqa
\label{eq:V2np}
&&m^4 V^{np}_2(\{\mu_k\})
= \nl
&&~~~ \dashint \prod_{j=3}^{4}  \left(
\frac{d \sigma_j}{\sigma_j+i \epsilon} \right)\! \int_{-1}^{1} \frac{d c_\theta}{2} \Phi^{(\mu_3,\mu_4)}_1[D^{np}], 
\eqa
where
$$D^{np} := D(\tau_1,\sigma_3+\mu_3,\tau_2,\sigma_4+\mu_4,\sigma^{34}_a,\sigma^{34}_b,\mu_1,\mu_2,\mu_0,\mu_5).$$ 
Now the loop momentum of the $m_3$ line  flows through two propagators of $D^{np}$. This provides an additional damping factor in \eqref{eq:V2np} with respect to \eqref{eq:V2D}, so that large $\sigma_{3,4}$ configurations do not lead to numerical inaccuracies and  no technical cut is required.
A few numerical results are collected in
Table \ref{tab:V2nl}.

\begin{figure}
\vskip -2.4cm
\hskip -4.3cm
\includegraphics[width=6.5in]{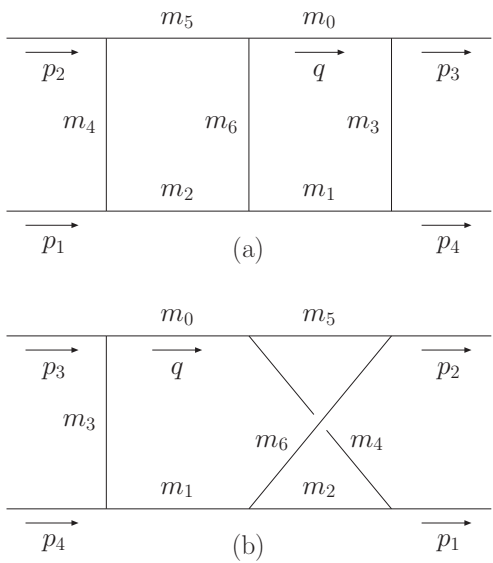}
\vskip -2.9cm
\caption{The planar (a) and non-planar (b) two-loop double boxes $B_2(\{\mu_k\})$ and $B^{np}_2(\{\mu_k\})$ of \eqref{eq:B2} and \eqref{eq:B2np} with masses $\{m^2_k/m^2\} =\{\mu_k\} :=  \{\mu_0,\mu_1,\mu_2,\mu_3,\mu_4,\mu_5\,\mu_6\}$.}
\label{fig:B2}   
\end{figure}

\begin{figure}
\vskip -4.3cm
\hskip -3.2cm
\includegraphics[width=5.7in]{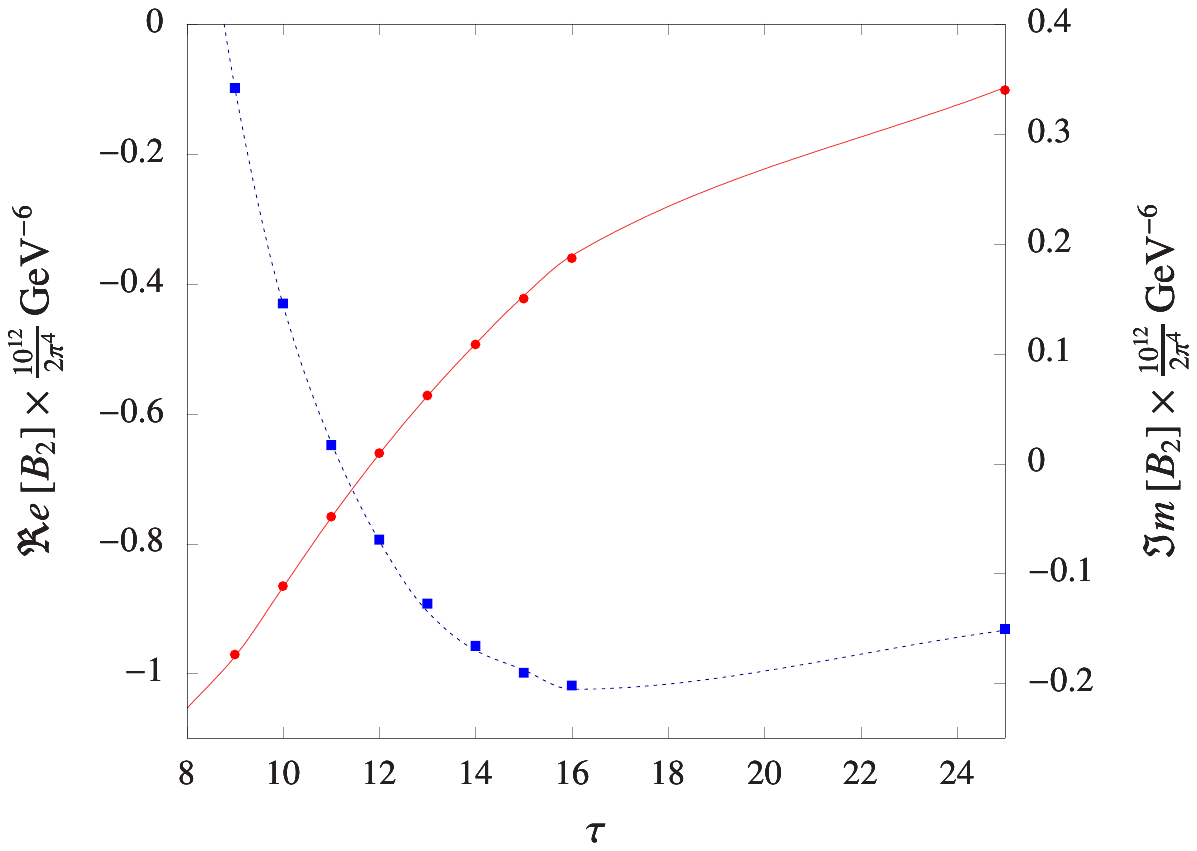}
\vskip -10.cm
\caption{The two-loop integral $B_2(\{\mu_k\})$ of \eqref{eq:B2}  as a function of $\tau$ with the input values listed in \eqref{eq:confB2}.
Red bullets (blue squares) refer to the real (imaginary) part obtained with $10^{9}$ MC shots per point using {\tt OneLOop} to compute $D_L$.
The solid and dashed lines are derived from Fig. 2 of \cite{Yuasa:2011ff}. 
}
\label{fig:B2plot}   
\end{figure}

\begin{figure}
\vskip -4.3cm
\hskip -3.2cm
\includegraphics[width=5.7in]{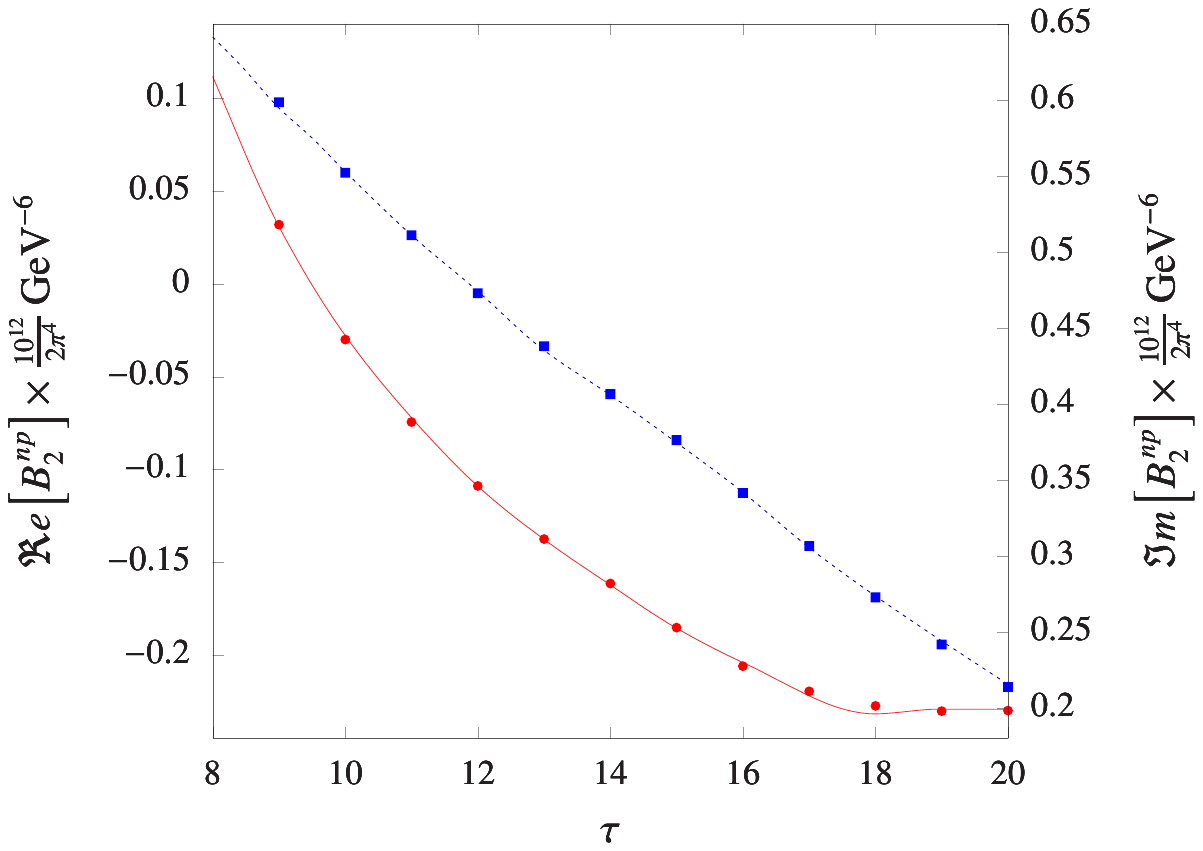}
\vskip -10.cm
\caption{The two-loop integral $B^{np}_2(\{\mu_k\})$ of \eqref{eq:B2np}  as a function of $\tau$ with the input values listed in \eqref{eq:confB2} (not $\{\mu_k\}= \{1,1,1,3.24,3.24,3.24,1\}$, as stated in the text of \cite{Yuasa:2011ff}).
Red bullets (blue squares) refer to the real (imaginary) part obtained with $10^{9}$ MC shots per point using {\tt OneLOop} to compute $D_R$.
The solid and dashed lines are derived from Fig. 3 of \cite{Yuasa:2011ff}. 
}
\label{fig:B2npplot}   
\end{figure}

\begin{table}
\caption{
The non-planar two-loop integral \eqref{eq:V2np} as a function of $\tau$ with equal internal masses $\{\mu_k\}= \{1,1,1,1,1,1\}$ and massless external momenta $\tau_1= \tau_2= 0$.
Numbers obtained with $10^{9}$ MC points by computing $D^{np}$ with {\tt OneLOop}. Statistical errors within parentheses.}
\label{tab:V2nl}       
\begin{tabular}{rl}
\hline\noalign{\smallskip}
 $\tau$ &  MC result \\
\noalign{\smallskip}\hline\noalign{\smallskip}
2.1 &   -1.538(8)$\times 10^{1}$ $-$$i$ 6(6)$\times 10^{-2}$ \\
10 &  8.7(1)$\times 10^{-1}$    $-$$i$ 2.5415(9)$\times 10^{1}$ \\
100 &   1.8848(6)    +$i$ 7.469(6)$\times 10^{-1}$  \\
1000 &  2.660(2)$\times 10^{-2}$   +$i$ 7.788(2)$\times 10^{-2}$  \\
\noalign{\smallskip}\hline
\end{tabular}
\end{table}
\paragraph{Planar and non-planar double box}
$~$ \vskip 5pt
\noindent The planar and non-planar two-loop double boxes are depicted in
Fig.~\ref{fig:B2} (a) and (b), respectively. They read
\bqa
\label{eq:B2}
B_2(\{\mu_k\})&=&\frac{1}{m^6}
\dashint \!\prod_{j=0}^{1}  \left(
\frac{d \sigma_j}{\sigma_j+i \epsilon} \right) \Phi^{(\mu_0,\mu_1,\mu_3)}_3[D_L], \\
\label{eq:B2np}
B^{np}_2(\{\mu_k\})&=&\frac{1}{m^6}
\dashint \! \prod_{j=0}^{1}  \left(
\frac{d \sigma_j}{\sigma_j+i \epsilon} \right) \Phi^{(\mu_0,\mu_1,\mu_3)}_3[D_R], 
\eqa
where
\bqa
D_{L} &:=& D(\tau_1,\tau_2,\sigma_0\!+\!\mu_0,\sigma_1\!+\!\mu_1,\tau,\sigma^{01}_b,\mu_2,\mu_4,\mu_5,\mu_6),\nl
D_{R} &:=& D(\tau_1,\sigma_0\!+\!\mu_0,\tau_2,\sigma_1\!+\!\mu_1,\sigma^{01}_a,\sigma^{01}_b,\mu_2,\mu_4,\mu_5,\mu_6), \nonumber
\eqa
are the one-loop boxes on the left and right sides of Fig.~\ref{fig:B2} (a) and (b), respectively. 
In Figs.~\ref{fig:B2plot} and \ref{fig:B2npplot} we present a comparison between our estimates and the results presented in \cite{Yuasa:2011ff} for the case
\bqa
\label{eq:confB2}
{\mu_k}&=& \{1,1,1,3.24,3.24,1,3.24\},~
\tau_i= 1,~\chi= -4, \nl
m&=& 50~{\rm GeV},
\eqa
in the region of $\tau$ where
$|\cos \theta_{13}| \le 1$. \footnote{An analytic continuation to nonphysical configurations is possible, although we did not try it.}
The agreement is very good. As benchmark values, we list in Table \ref{tab:B2s} the MC entries of Figs.~\ref{fig:B2plot} and \ref{fig:B2npplot}, together with their statistical errors.   

\begin{figure}[t]
\vskip -4.5cm
\hskip -3.9cm
\includegraphics[width=6.2in]{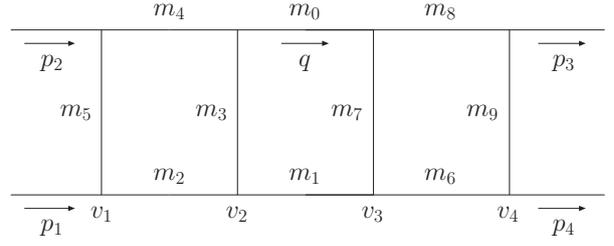}
\vskip -4.2cm
\caption{The three-loop planar box $B_3$ of \eqref{eq:B3}.}
\label{fig:B3}   
\end{figure}

\begin{figure}
\vskip -4.5cm
\hskip -3.9cm
\includegraphics[width=6.2in]{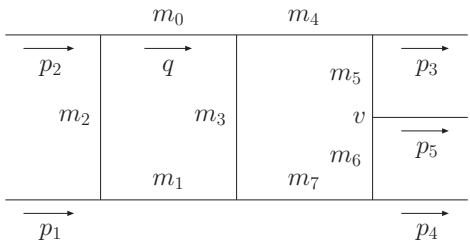}
\vskip -4.2cm
\caption{The two-loop planar pentabox $E_2$ of \eqref{eq:E2}.}
\label{fig:E2}   
\end{figure}

\begin{table}
\caption{The MC entries of Figs.~\ref{fig:B2plot} and \ref{fig:B2npplot} in units of ${\rm GeV}^{-6}$. The MC errors are indicated between parentheses.}
\label{tab:B2s}       
\begin{tabular}{rll}
\hline\noalign{\smallskip}
 $\tau$   & $10^{12} B_2/(2 \pi^4)$ & $10^{12} B^{np}_2/(2 \pi^4)$ \\
\noalign{\smallskip}\hline\noalign{\smallskip}
9  &  -0.971(1)  +$i$ 0.342(1)    &   0.0318(9) +$i$  0.5985(9) \\
10 &  -0.8655(8) +$i$ 0.1457(8)   &  -0.0301(7) +$i$  0.5521(7) \\
11 &  -0.7586(6) +$i$ 0.0168(6)   &  -0.0746(6) +$i$  0.5110(5) \\ 
12 &  -0.6607(7) $-$$i$ 0.0692(6) &  -0.1090(5) +$i$  0.4729(5) \\
13 &  -0.5720(5) $-$$i$ 0.1275(5) &  -0.1376(4) +$i$  0.4380(4) \\
14 &  -0.4934(4) $-$$i$ 0.1659(4) & -0.1616(3)  +$i$  0.4065(3) \\
15 &  -0.4229(3) $-$$i$ 0.1903(3) &  -0.1853(3) +$i$  0.3763(3) \\
16 &  -0.3607(3) $-$$i$ 0.2019(3) &  -0.2061(3) +$i$  0.3414(2) \\
17 &  ~~~---~~~&  -0.2196(2)+$i$  0.3065(2) \\
18 &  ~~~---~~~& -0.2275(2) +$i$  0.2728(2) \\
19 &  ~~~---~~~& -0.2303(2) +$i$  0.2418(2) \\
20 &  ~~~---~~~& -0.2301(2) +$i$  0.2138(2) \\
25 &  -0.1017(1) $-$$i$ 0.1506(1) & ~~~---~~~\\
\noalign{\smallskip}\hline
\end{tabular}
\end{table}
\paragraph{More complex structures}
$~$ \vskip 5pt
\noindent In all cases presented so far we could perform an easy analytic integration over at least one angular variable of the loop momentum $\omega$. When the number of loops and legs increases, integrating analytically over $c_\theta$ and/or $\phi$ is not trivial any more, so that the number of numerical integrations required by the gluing procedure reaches its maximum value, i.e. four.
In this paragraph we give a few  examples of the gluing approach in these more complex situations. In particular, we use the multichannel approach of Sect. \ref{sec:method1} to  study the scalar triple box $B_3$ and two-loop pentabox $E_2$ of Figs.~\ref{fig:B3} and \ref{fig:E2}, respectively.

As for $B_3$, inserting \eqref{eq:one} into the integrand gives \bqa
\label{eq:B3}
m^8 B_3&=& \dashint \prod_{j=0}^{1}  \left(
\frac{d \sigma_j}{\sigma_j+i \epsilon} \right)
\int_{-1}^1 \frac{d c_\theta}{2} \int_0^{2 \pi} \frac{d \phi}{2 \pi} \nl
&&\times \Phi_1^{(\mu_0,\mu_1)}
\left[D_L(c_\theta) D_R(c_\theta,s_\phi)\right],
\eqa
where
\bqa
D_L(c_\theta) &:=& D(\sigma_1+\mu_1,\sigma_0+\mu_0,\tau_2,\tau_1,\tau,\sigma^{01}_b,\nl
&&\mu_2,\mu_3,\mu_4,\mu_5), \nl
D_R(c_\theta,s_\phi) &:=& D(\sigma_1+\mu_1,\sigma_0+\mu_0,\tau_3,\tau_4,\tau,\sigma^{01}_c,\nl
&&\mu_6,\mu_7,\mu_8,\mu_9), 
\eqa
are the one-loop boxes on the left and right sides of  Fig.~\ref{fig:B3}.
The change of variable 
\bqa
\label{eq:phitoy}
y= c_\phi
\eqa
produces 
\bqa
\label{eq:B3a}
m^8 B_3 &=& \frac{1}{8 \tau}
\dashint \prod_{j=0}^{1}  \left(
\frac{d \sigma_j}{\sigma_j+i \epsilon} \right) \lambda^{\frac{1}{2}} \theta(\lambda)
\int_{-1}^1 d c_\theta\, D_L(c_\theta)
\nl
&& \hskip -0.8cm \times \int_{-1}^1 \frac{dy}{\sqrt{1-y^2}} \sum_{n=0}^1  
D_R\big(c_\theta,(-1)^n\sqrt{1-y^2}\big).
\eqa
When
$\tau= 50$, $\chi= -4$, $\tau_i= \mu_0= \mu_1= 1$, $\mu_{\tiny 2\div 9}= 0$,
\eqref{eq:B3a} gives, with $10^9$ MC shots
and using {\tt OneLOop} to evaluate $D_{L,R}$, 
\bqa
\label{eq:B3number}
m^8 B_3 = -9.393(9) -i\, 0.374(9).
\eqa
Note that it is not difficult to deal with non-planar configurations. For instance, the
diagram obtained from Fig.~\ref{fig:B3} by interchanging the vertices $v_1 \leftrightarrow v_2$ and 
$v_3 \leftrightarrow v_4$ is as in \eqref{eq:B3}, but with
\bqa
D_L(c_\theta) &:=& D(\tau_1,\sigma_0+\mu_0,\tau_2,\sigma_1+\mu_1,\sigma^{01}_a,\sigma^{01}_b, \nl
&&\mu_2,\mu_3,\mu_4,\mu_5), \nl
D_R(c_\theta,s_\phi) &:=& D(\tau_4,\sigma_0+\mu_0,\tau_3,\sigma_1+\mu_1,\sigma^{01}_d,\sigma^{01}_c,\nl
&&\mu_6,\mu_7,\mu_8,\mu_9).
\eqa

Finally, the pentabox depicted in Fig.~\ref{fig:E2} reads
\bqa
\label{eq:E2}
m^8 E_2&=& \dashint \prod_{j=0}^{1}  \left(
\frac{d \sigma_j}{\sigma_j+i \epsilon} \right)
\int_{-1}^1 \frac{d c_\theta}{2} \int_0^{2 \pi} \frac{d\phi}{2 \pi} \nl
&& \times \Phi_1^{(\mu_0,\mu_1)}
\left[
\frac{E(\omega)}{\sigma_2+i \epsilon}
\right],
\eqa
where $E(\omega)$ is the one-loop pentagon on the right side, which depends upon ten independent invariants\,--\,built from the momenta $q\!=\! m \omega, \,p_{3\div 5}$\,--\,and the five masses $m_{3\div 7}$.
The presence of the $1/(\sigma_2+i \epsilon)$ propagator forces one to trade the integral over $c_\theta$ for an integration over $\sigma_2$ to be dealt with the method of Sect. \ref{sec:2}. This, together with the change of variable 
of \eqref{eq:phitoy}, gives
\bqa
\label{eq:E2a}
m^8 E_2&=& \frac{1}{4 \tau k}
\dashint \prod_{j=0}^{2}  \left(
\frac{d \sigma_j}{\sigma_j+i \epsilon} \right)
\theta(\lambda)\, \theta(1-|c_\theta|) \nl
&&\times \int_{-1}^1 \frac{d y}{\sqrt{1-y^2}} \sum_{n=0}^1 E(\omega_n),
\eqa
where
\bqa
\label{eq:sigmat}
k \lambda^{\frac{1}{2}} c_\theta&=&
\sigma_0+\mu_0+\sigma_1+\mu_1 -2(\sigma_2+\mu_2-\tau_2)-\tau \nl
&&+ \sigma_\tau \frac{\tau_{12}}{\tau}, \nl
\omega_{0,1} &:=& 
\frac{1}{2 \sqrt{\tau}}\left(
\sigma_\tau, \lambda^{\frac{1}{2}} c_\theta, 
\pm \lambda^{\frac{1}{2}} s_\theta \sqrt{1-y^2},
\pm \lambda^{\frac{1}{2}}s_\theta y
\right), \nl
\sigma_\tau &:=& \tau +\sigma_0+\mu_0-\sigma_1-\mu_1. 
\eqa
To provide a benchmark value, we have chosen to evaluate \eqref{eq:E2a} with $\mu_{0\div 2}= 1$, $\mu_{3\div 7}= 0$
at a particular $p_1+p_2 \to p_3+p_4+p_5$ phase-space point satisfying $\tau= 50$, 
$\tau_1= \tau_2=0$ and $\tau_{3 \div 5}= 1$, namely
\bqa
\frac{p_1}{m}&=& \frac{\sqrt{\tau}}{2}\left(1,1,0,0\right), \hskip 0.5cm 
\frac{p_2}{m}= \frac{\sqrt{\tau}}{2}\left(1,-1,0,0\right), \nl
\frac{p_3}{m}&=&  \frac{3}{8} \sqrt{\tau} \left(1,r,r,0\right), \nl
\frac{p_4}{m}&=& \frac{3}{8} \sqrt{\tau}\left[
1,
r \left(c_{34}-\frac{s_{34}}{\sqrt{2}}\right),
r \left(c_{34}+\frac{s_{34}}{\sqrt{2}}\right),
r s_{34}
\right], \nonumber
\eqa
with $r\!=\! \sqrt{193/450}$, $c_{34}\!=\! -159/193$ and $s_{34}\!=\! \sqrt{1-c^2_{34}}$.
We employed {\tt CutTools} \cite{Ossola:2007ax} to reduce $E(\omega_n)$ to one-loop boxes. Computing the latter with
$\tt{OneLOop}$ gives, with $10^9$ MC shots,
\bqa
\label{eq:E2number}
m^8 E_2=  0.1125(1)-i\,0.0041(1).
\eqa
Once again, non-planar configurations are obtained without extra effort. For instance, if the vertex $v$ of Fig.~\ref{fig:E2} is moved to the $m_3$ line, \eqref{eq:E2} still holds by simply modifying the one-loop pentagon $E(\omega)$ accordingly.

\section{UV divergences}
\label{sec:5}
\begin{figure}
\vskip -4.6cm
\hskip -5.3cm
\includegraphics[width=6.5in]{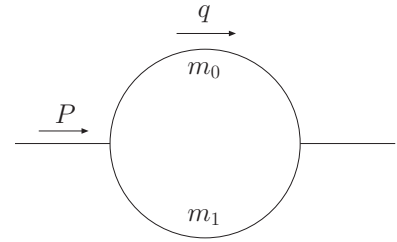}
\vskip -4.8cm
\caption{The one-loop self-energy diagram of 
\eqref{eq:BFDR0}.}
\label{fig:BFDR}   
\end{figure}

We deal with ultraviolet divergent integrals following the FDR approach of \cite{Pittau:2012zd}, in which the divergent configurations are extracted from the original integrand by partial fractioning. \footnote{At one loop, FDR is equivalent to the $\overline{\rm MS}$ scheme of Dimensional Regularization.} The resulting expressions are integrable in four dimensions and nicely match the algorithm of Sec.~\ref{sec:2}. 
We illustrate our procedure by means of the scalar two-point function of Fig.~\ref{fig:BFDR},
\bqa
\label{eq:BFDR0}
B_{\scriptscriptstyle \rm FDR} = \int [d^4q] \frac{1}{\bar D_0 \bar D_1},
\eqa
where
\bqa
\label{eq:den0}
\bar D_0 := \bar q^2 -m_0^2,~~\bar D_1 := \bar q^2-M^2_1(q),
\eqa
with
$M^2_1(q) := m^2_1+2 (q \cdot P)-P^2$ and $\bar q^2 := q^2-\mu^2+i\epsilon$.
By partial fractioning  \footnote{The UV divergent piece is dubbed {\em vacuum} and written between square brackets by convention.}
\bqa
\frac{1}{\bar D_0 \bar D_1}=
\left[\frac{1}{\bar q^4}\right]
+\frac{m_0^2}{\bar q^2 \bar D_0 \bar D_1}
+\frac{M^2_1(q)}{\bar q ^4 \bar D_1}.
\eqa
This gives, by definition of FDR integration,
\bqa
B_{\scriptscriptstyle \rm FDR} := \lim_{\mu \to 0}\int d^4q
\left.\left(
\frac{m_0^2}{\bar q^2 \bar D_0 \bar D_1}
+\frac{M^2_1(q)}{\bar q ^4 \bar D_1}
\right)\right|_{\mu = \mur},
\eqa
where $\mur$ is the finite renormalization scale.
It is convenient to take the limit ${\mu \to 0}$ directly at the integrand level and substitute $\mu$
with $\mur$ only in the logarithms. This is achieved by rewriting
\bqa
\label{eq:BFDR2}
B_{\scriptscriptstyle \rm FDR} = \int d^4q
\left\{
\frac{1}{D_0 D_1}- \frac{1}{(q^2-\mur^2+i \epsilon)^2}
\right\},
\eqa
where $D_i := \bar D_i+\mu^2$. By doing so, $\mu^2$ is dropped everywhere, except in the logarithmically UV divergent part of the {\em vacuum}, where it is replaced by $\mur^2$. In this way, no $\mu \to 0$ limit is required, so that \eqref{eq:BFDR2} is a good starting point for a numerical treatment. In what follows we describe how 
the methods of Sects.~\ref{sec:3} and \ref{sec:4} can be adapted to deal with \eqref{eq:BFDR2}. More complex multi-loop UV divergent configurations can be treated likewise.
\subsection{Integrating over the loop energy component}
We take $m_1=m_0=m$ for simplicity.
A rescaling as in \eqref{eq:omegaq} produces
\bqa
\label{eq:BFDR1}
B_{\scriptscriptstyle \rm FDR} = 4 \pi
\int_0^\infty\! \rho^2 d \rho\,I_t, 
\eqa
where
\bqa
I_t &:=& \int_{- \infty}^{+ \infty} dt \left \{
\frac{1}{(t^2-R^2_0+i \epsilon)((t-\sqrt{\tau})^2-R^2_0+i \epsilon)} \right.\nl
&&\left.-\frac{1}{(t^2-R^2_\nu+i \epsilon)^2} \right\},
\eqa
with 
\bqa
\label{eq:nuscale}
R^2_0 := {\rho^2+1},~~ R^2_\nu := {\rho^2+\nu},~~\nu := {\mur^2}/{m^2}.
\eqa
The Cauchy integral theorem allows one to compute
\bqa
I_t&=& \frac{i\pi}{2}  \left\{\frac{1}{\big(R_0-i \epsilon\big)\big( R_0-\sqrt{\tau}/2 -i \epsilon\big)\big(R_0+\sqrt{\tau}/2 -i \epsilon\big)} \right. \nl
&& \left. -\frac{1}{(R_\nu-i \epsilon)^3} \right\}.
\eqa
Inserting this in \eqref{eq:BFDR1} and using $r= R_0-\sqrt{\tau}/2$ as a new integration variable gives
\bqa
\label{eq:BFDR3}
B_{\scriptscriptstyle \rm FDR} = 2 i \pi^2
\dashint dr \frac{F(r)}{r-i \epsilon},
\eqa
where
\bqa
F(r) &=& \theta\big(r-1+\sqrt{\tau}/2\big) \sqrt{\big(r+\sqrt{\tau}/2\big)^2-1} \\
&&\times 
\left\{
\frac{1}{r+\sqrt{\tau}}
-r \frac{r+\sqrt{\tau}/2}{\big[\big(r+\sqrt{\tau}/2\big)^2-1+\nu\big]^{3/2}}
\right\}. \nonumber
\eqa
If $\sqrt{\tau} > 2$ the pole at $r= i \epsilon$  migrates towards the integration contour when $\epsilon \to 0$. Treating this with our numerical approach produces the results collected in  Table \ref{tab:BFDR}.

Integrals with a polynomial degree of divergence can be treated in exactly the same way. As an example, \ref{app:b} details the case of the  one-point function
\bqa
\label{eq:AFDR0}
A_{\scriptscriptstyle \rm FDR} = \int [d^4q] \frac{1}{\bar D_0}.
\eqa

\begin{table*}
\caption{The integral of \eqref{eq:BFDR0} with  $m=m_0=m_1=2 \mur$ as a function of $\sqrt{\tau}$. The numerical estimates are computed by sampling \eqref{eq:BFDR3}
with $10^8$ MC shots and the analytic results are obtained with ${\tt OneLOop}$. MC errors between parentheses.}
\label{tab:BFDR}       
\begin{tabular}{rll}
\hline\noalign{\smallskip}
 $\sqrt{\tau}$   & Numerical result & Analytic result \\
\noalign{\smallskip}\hline\noalign{\smallskip}
0.2 &  2(2)$\times 10^{-6}$  $-$$i$  1.3614(3)$\times 10^{1}$ &  0 $-$$i$  1.3616$\times 10^{1}$                     \\
0.4 &  2(2)$\times 10^{-6}$  $-$$i$  1.3411(3)$\times 10^{1}$ &  0 $-$$i$  1.3415$\times 10^{1}$                     \\
0.6 &  2(2)$\times 10^{-6}$  $-$$i$  1.3065(3)$\times 10^{1}$ &  0 $-$$i$  1.3068$\times 10^{1}$                     \\ 
1.5 &  2(2)$\times 10^{-6}$  $-$$i$  8.705(1)                 &  0 $-$$i$  8.7063                                   \\
1.9 &  4(4)$\times 10^{-7}$  $-$$i$  2.0737(3)                &  0 $-$$i$  2.0739                                   \\
2.1 & -9.455(1)  +$i$  4.162(1)                              & -9.4541  +$i$  4.1616                               \\
4   & -2.6854(3)$\times 10^{1}$ $-$$i$ 1.6453(5)$\times 10^{1}$ & -2.6852$\times 10^{1}$ $-$$i$  1.6456$\times 10^{1}$  \\
10  & -3.0377(5)$\times 10^{1}$ $-$$i$  3.828(2)$\times 10^{1}$ & -3.0380$\times 10^{1}$ $-$$i$  3.8280$\times 10^{1}$  \\
50  & -3.0990(6)$\times 10^{1}$ $-$$i$  7.112(5)$\times 10^{1}$ & -3.0981$\times 10^{1}$ $-$$i$  7.1094$\times 10^{1}$  \\
100 & -3.1009(6)$\times 10^{1}$ $-$$i$  8.473(7)$\times 10^{1}$ & -3.1000$\times 10^{1}$ $-$$i$  8.4825$\times 10^{1}$  \\
\noalign{\smallskip}\hline
\end{tabular}
\end{table*}

\subsection{Gluing substructures}
The gluing approach of Sec.~\ref{sec:4} can be easily extended to \eqref{eq:BFDR0}.
Inserting \eqref{eq:one} in \eqref{eq:BFDR2} gives
\bqa
\label{eq:BFDRg}
B_{\scriptscriptstyle \rm FDR}(\nu)&=& \frac{\pi}{2 \tau}
 \dashint \prod_{j=0}^{1}  \left(
\frac{d \sigma_j}{\sigma_j+i \epsilon}
\right)  \lambda^{\frac{1}{2}} \theta(\lambda)\, J(\nu),
\eqa
with
\bqa
J(\nu) &=& 1-\frac{\sigma_0 \sigma_1}{(\sigma_0+\mu_0-\nu+i \epsilon)^2} \eqa 
and $\nu$ defined in \eqref{eq:nuscale}.
The presence of a double pole is an obstacle to a direct numerical treatment of \eqref{eq:BFDRg}. In fact, our algorithm is designed to deal with single poles only.
However, we observe that, if $\nu$ has a finite imaginary part, the singularity never approaches the real axis. In particular, $B_{\scriptscriptstyle \rm FDR}(-i\nu)$ is better suited than $B_{\scriptscriptstyle \rm FDR}(\nu)$ to be evaluated numerically if $\nu \in {\Re}$. Besides,
the connection between the two can be derived by differentiating \eqref{eq:BFDR2},
\bqa
\frac{\partial B_{\scriptscriptstyle \rm FDR}}{\partial \mur^2}= -2 \int \frac{d^4q}{(q^2-\mur^2+i \epsilon)^3}= \frac{i \pi^2}{\mur^2},
\eqa
which gives
\bqa
\label{eq:BFDRg0}
B_{\scriptscriptstyle \rm FDR}(\nu)= B_{\scriptscriptstyle \rm FDR}(-i\nu^\prime) -\frac{\pi^3}{2}
- i \pi^2 \ln \frac{\nu^\prime}{\nu}.
\eqa
$B_{\scriptscriptstyle \rm FDR}(-i\nu^\prime)$ still suffers from
numerical inaccuracies of type \eqref{eq:ab} (a). To cure this, we locally subtract from it an approximant, $
\tilde B_{\scriptscriptstyle \rm FDR}(-i\nu^\prime) = 0$,
constructed in such a way that, after changing variables as in \eqref{eq:s01}, all  cuts  and  poles  lie  in  the  lower $\sigma_0 $ complex half-plane. This is obtained by replacing in \eqref{eq:BFDRg}
$\lambda^{\frac{1}{2}}\theta(\lambda) \to \sqrt{\lambda^2-4\tau i \epsilon}\, \theta(\lambda+4\tau\sigma_0)$. In summary, we rewrite
\bqa
\label{eq:BFDRgp}
B_{\scriptscriptstyle \rm FDR}(-i\nu^\prime) &=& 
\frac{\pi}{2 \tau}
 \dashint \prod_{j=0}^{1}  \left(
\frac{d \sigma_j}{\sigma_j+i \epsilon}
\right) 
 \\
&&\hspace{-1.cm} \times  
 \big[\lambda^{\frac{1}{2}} \theta(\lambda)-
\sqrt{\lambda^2-4\tau i \epsilon}\, \theta(\lambda+4\tau\sigma_0)\big]
J(-i\nu^\prime). \nonumber
\eqa

In Table \ref{tab:BFDRg} we present our estimates for $B_{\scriptscriptstyle \rm FDR}(1/4)$ with $\mu_0=\mu_1=1$ obtained by means of Eqs. \eqref{eq:BFDRg0} and \eqref{eq:BFDRgp}. The figures match the results of Table \ref{tab:BFDR}, although with larger errors. However, we point out that the gluing method is more flexible when it comes to generic kinematics. For instance, with
$\tau= -10$, $\mu_0=1$, $\mu_1=4$, one obtains, with $10^8$ MC points,
\bqa
B_{\scriptscriptstyle \rm FDR}(1/4)= 
0.02(2) -i\, 27.20(3),
\eqa
to be compared to the analytic value $-i\, 27.220$.

\begin{table}
\caption{The integral of \eqref{eq:BFDRg0}
with $\nu= 1/4$ and $\mu_0=\mu_1= 1$. The estimates are obtained by sampling 
with $10^8$ MC shots \eqref{eq:BFDRgp} evaluated at $\nu^\prime=1$. MC errors between parentheses.}
\label{tab:BFDRg}       
\begin{tabular}{rl}
\hline\noalign{\smallskip}
 $\sqrt{\tau}$   & $\displaystyle B_{\scriptscriptstyle \rm FDR}(-i\nu^\prime)  - \frac{\pi^3}{2}
  - i \pi^2 \ln \frac{\nu^\prime}{\nu}$  \\
\noalign{\smallskip}\hline\noalign{\smallskip}
0.2 &  -4(4)$\times 10^{-3}$  $-$$i$  1.357(3)$\times 10^{1}$ \\
0.4 &  -3(3)$\times 10^{-2}$  $-$$i$  1.342(2)$\times 10^{1}$ \\
0.6 &  -1(1)$\times 10^{-2}$  $-$$i$  1.308(2)$\times 10^{1}$ \\ 
1.5 &  -1(1)$\times 10^{-2}$  $-$$i$  8.69(2)                 \\
1.9 &   2(2)$\times 10^{-3}$  $-$$i$  2.04(2)                \\
2.1 & -9.45(1)  +$i$  4.19(2)                               \\
4   & -2.688(2)$\times 10^{1}$ $-$$i$ 1.655(3)$\times 10^{1}$ \\
10  & -3.039(4)$\times 10^{1}$ $-$$i$  3.825(5)$\times 10^{1}$ \\
50  & -3.100(8)$\times 10^{1}$ $-$$i$  7.12(1)$\times 10^{1}$ \\
100 & -3.09(1)$\times 10^{1}$ $-$$i$  8.49(2)$\times 10^{1}$ \\
\noalign{\smallskip}\hline
\end{tabular}
\end{table}

\section{IR divergences}
\label{sec:6}
We deal with IR divergent integrals by means of the FDR approach of 
\cite{Pittau:2013qla}, where a small mass $\mu$, added to judiciously chosen propagators, is used as a regulator of both infrared and collinear divergences. 
In this section, we illustrate how this allows one to combine virtual and real contributions prior to integration. After that, our method can be used to evaluate numerically loop integrals where the IR configurations are locally subtracted. We study, in particular, the IR divergent scalar triangle
\bqa
\label{eq:C0IR}
C_{\rm IR} &:=& \lim_{\mu \to 0}
\int d^4q 
\frac{1}{D_0 D_1 D_2}, \\
 D_0 &:=& q^2-\mu^2+i \epsilon, \nl
 D_1 &:=& (q+p_1)^2-\mu^2+i\epsilon,~D_2 := (q-p_2)^2-\mu^2+i \epsilon,\nonumber
\eqa
that appears in a $P \to p_1+p_2$ decay with $p_i^2= 0$ and $s:= P^2$. However, our findings can be generalized to more complex environments.

\begin{figure}
\vskip -4.8cm
\hskip -4cm
\includegraphics[width=6.5in]{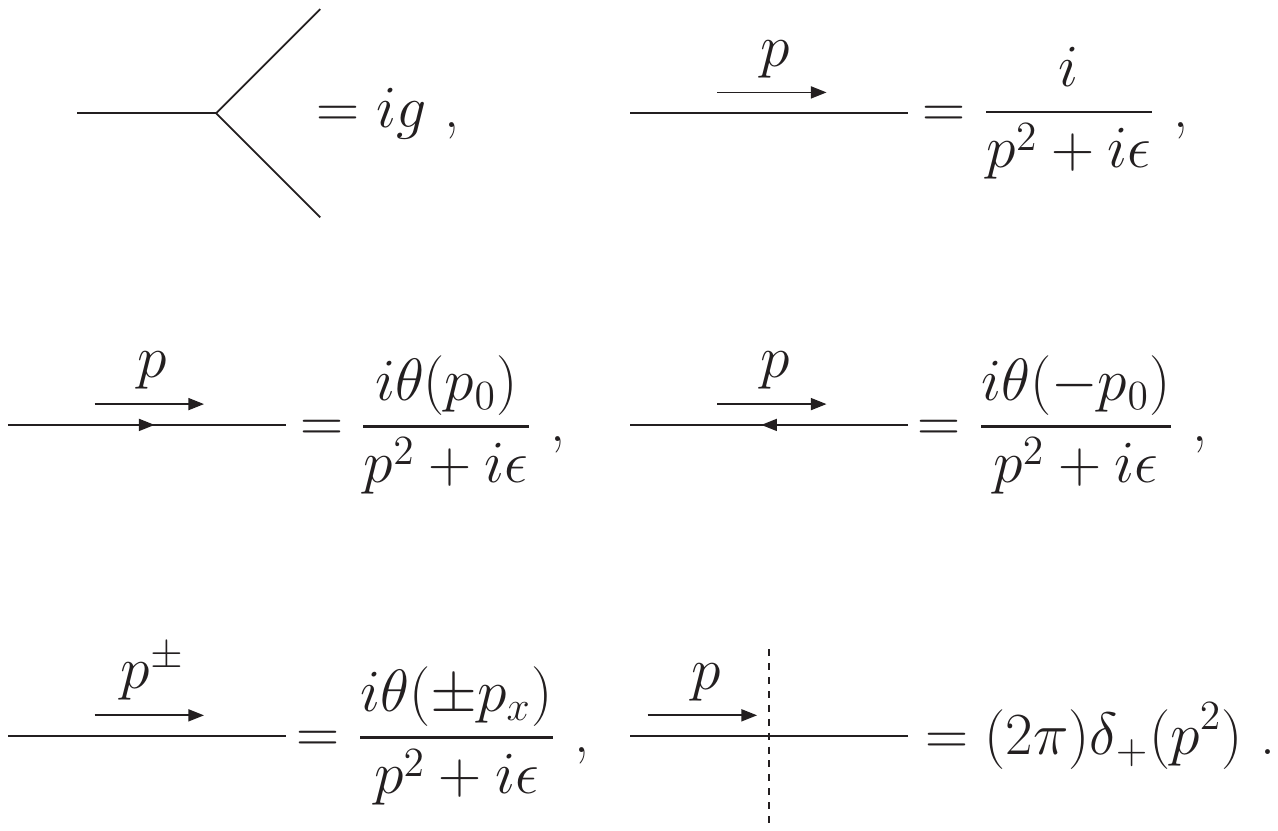}
\vskip -2.5cm
\caption{The Feynman rules of the $g \varphi^3$ theory. A special notation is used for propagators with positive and negative values of 
$p_0$ and $p_x$, which, in our convention, coincides with the direction of the back-to-back final-state particles in the $P$ rest-frame. The complex conjugate of such rules is used in the r.h.s.  of diagrams cut by a dashed line.}
\label{fig:figFR}   
\end{figure}

\begin{figure}
\vskip -0.7cm
\hskip -3.7cm
\includegraphics[width=6.4in]{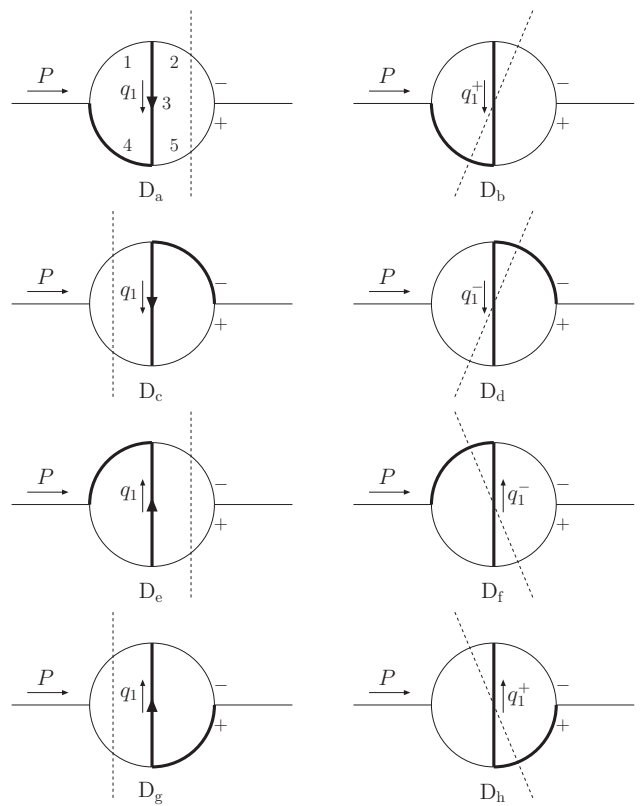}
\vskip -1cm
\caption{The two- and three-particle cuts contributing to $\varphi^{\ast} \to \varphi \varphi (\varphi)$ in the $P$ 
rest frame where the propagators 2 and 5 have negative and positive components of the momentum along $x$, as indicated by the $\mp$ labels attached to them. Thick lines represent the propagators to which $\mu^2 \to 0$ is added, as explained in the text.}
\label{fig:cuttings}   
\end{figure}

Our strategy is based on combining together cut-diagrams that are individually divergent, but whose sum is finite. We use a scalar massless $g \varphi^3$ theory defined through the Feynman rules of Fig.~\ref{fig:figFR},
where we have introduced propagators with positive and negative values of the energy and the momentum component along the $x$ direction. The cuts contributing to $\varphi^{\ast} \to \varphi \varphi (\varphi)$ are listed in Fig.~\ref{fig:cuttings} where, to make contact with \eqref{eq:C0IR},
\bqa
\label{eq:fromDtoC}
{\rm D_{a}+D_{e}} &=&  4 \pi^2 g^4 \frac{\pi}{2}
(i\, C_{\rm IR}), \nl
{\rm D_{c}+D_{g}} &=&  4 \pi^2 g^4 \frac{\pi}{2}
(i\, C_{\rm IR})^\ast.
\eqa
The diagrams are organized in pairs sharing collinear singularities. For instance, in $\rm D_{a}$ the energy component of propagator 1 is the sum of those of propagators 2 and 3. Thus, propagators 1 and 3 never pinch in the $q^0_{1}$ complex plane, and propagator 3 can only become collinear to 4. Likewise, in $\rm D_{b}$ the sign of the momentum components along $x$ only allows 
particles 3 and 4 to become collinear to 5. In both cases, we regulate the singular splitting by including a small mass $\mu$ in propagators 3 and 4, leaving 5 massless. \footnote{Note that adding $\mu^2$ also to 1 and/or 2 does not change the asymptotic $\mu \to 0$ limit of the result. This is used, for instance, in \eqref{eq:fromDtoC}.}
In summary, $\rm D_{a}+D_{b}$ is free of collinear divergences, and the same happens for $\rm D_{c}+D_{d}$. In addition,  $\rm D_{a}+D_{b}+D_{c}+D_{d}$ is also free of infrared singularities.
A similar reasoning applies to $\rm D_{e}+D_{f}+D_{g}+D_{h}$, but with an opposite sign of the energy component of $q_1$.
The previous analysis shows that the three-particle cuts
$\rm D_{b}$ and $\rm D_{d}$ can be used as local countertems for
$\rm D_{a}+D_{c}$. \footnote{An analogous procedure holds for the last four cuts of Fig.~\ref{fig:cuttings}.} This requires  common reference frames. One can employ two different routings for   $\rm D_{a}+D_{b}$ and $\rm D_{c}+D_{c}$. However, they must coincide in the limit $q_1 \to 0$ to guarantee the cancellation of the soft behaviour of  $\rm D_{a}+D_{b}+D_{c}+D_{d}$.
In particular, when computing $\rm D_{a,b}$ we assign a momentum $q_2$ to propagator 4, from left to right, and choose
\bqa
\label{eq:routing1}
\omega_1 &:=& {q_1}/{\sqrt{s}}= (t_1,\rho_1,0,0), \nl
\omega_2 &:=& {q_2}/{\sqrt{s}}= 
(t_2,\rho_2 c_\theta, \rho_2 s_\theta s_\phi, s_\theta c_\phi).
\eqa
On the other hand, we calculate $\rm D_{c,d}$
with $q_2$ assigned to propagator 5 and
\bqa
\label{eq:routing2}
\omega_1 &=& (t_1,-\rho_1,0,0), \nl
\omega_2 &=& 
(t_2,\rho_2 c_\theta, \rho_2 s_\theta s_\phi, s_\theta c_\phi).
\eqa
The result of the computation is reported in \ref{app:c} in terms of integrals over 
\bqa
\label{eq:eta}
R_i := \sqrt{\rho_i^2+\eta},~{\rm with}~\eta:= \mu^2/s.
\eqa
It is convenient to further split $\rm D_{a,c}^{\phantom s}=  D^{s}_{a,c}+D^{u}_{a,c}$, where the superscripts $\!\phantom{\!.}^{\rm s,u}$ refer to the
subtracted and unsubtracted regions, which correspond to the integration intervals $\sqrt{\eta} < R_1 < 1/2$ and $1/2 < R_1 < \infty$, respectively. In fact, $\rm D_{b,d}$
contribute in the subtracted region only, and $\rm D^{u}_{a}+D^{u}_{c}$ is free of IR singularities,
\bqa
\label{eq:IRuns}
\frac{s}{g^4} ({\rm D^{u}_{a}+D^{u}_{c}})&=&
-2 \pi^5 \big(\ln^2 2+2 \rm{Li}_2 (-1/2)\big) \nl
&=& 254.838137\cdots.
\eqa
An analytic calculation \cite{Pittau:2013qla} shows that
$\displaystyle \rm \sum_{j=a,b,c,d} D_j= 0$. Hence, one must have
\bqa
\label{eq:KIR}
K :=\! \frac{s}{g^4}\big({\rm D^{s}_{a}+D^{s}_{c}+D_{b}+D_{d}}\big)=\! -254.838137\cdots.
\eqa
In Table \ref{tab:KIR} we display our numerical estimate of $K$ based on Eqs. \eqref{eq:Db1}, \eqref{eq:Da1} and  \eqref{eq:Dd1}. The correct result is precisely approached and the MC error does not grow when decreasing $\eta$, which is an indication that the local cancellation works as expected. 
Finally, we point out that the outlined strategy can be turned into a fully exclusive
local subtraction algorithm by introducing suitable phase-space mappings, as described in \cite{Gnendiger:2017pys}.

\begin{table}
\caption{The combination of cut-diagrams defined in \eqref{eq:KIR} as a function of $\eta$ in \eqref{eq:eta}. Numbers obtained with 
$10^{10}$ MC shots.}
\label{tab:KIR} 
\begin{tabular}{cc}
\hline\noalign{\smallskip}
 $\eta$ &  $K$\\
\noalign{\smallskip}\hline\noalign{\smallskip}
$10^{-10}$ & -254.81(1)    \\
$10^{-11}$ & -254.83(1)    \\
$10^{-12}$ & -254.84(1)    \\
\noalign{\smallskip}\hline
\end{tabular}
\end{table}

\section{Conclusion and outlook}
\label{sec:7}
We have presented a flexible method for the numerical treatment of loop integrals in four-dimensional Minkowski space, without the need of explicit contour deformation. This is achieved  by exploiting the $i\epsilon$ prescription with a small finite value of $\epsilon$ and making changes of variables to reduce the variance of both the real and imaginary parts of the integrand. We propose a semi-numerical approach, in which an analytic integration over loop time-components is followed by multichannel Monte Carlo integration.  In some cases, further integrations can be performed before the final numerical step. The method lends itself readily to the evaluation of complex multi-loop structures by gluing together simpler substructures.  It also deals easily with processes involving many different external and propagator mass scales, where analytical results are difficult to obtain.

In practice, we find that $10^9$ Monte Carlo shots with $\epsilon\sim 10^{-7}$ (in terms of some relevant mass scale) can yield relative precision of the order of $10^{-4}$ for one-loop diagrams and $10^{-3}$ for two- and three-loops obtained by gluing together analytical results for one-loop substructures.
{As for the performance of our algorithms, we report in Table \ref{tab:timings} the time to produce $10^6$ MC shots with method 1 for a few representative cases. It ranges from a few tenths of a second to more than a minute.
Method 2 gives somewhat slower timings.}
\begin{table}
\caption{{Time to generate $10^6$ MC shots on a single 2.2 GHz processor.}}
\label{tab:timings} 
\begin{tabular}{ccc}
\hline\noalign{\smallskip}
{Type of integral} &{Location} & {Time [s]} \\
\noalign{\smallskip}\hline\noalign{\smallskip}
 {One-loop triangle} & {Last row of Table \ref{tab:5c}} & {0.25} \\
 {Two-loop self-energy} & {Eq. \eqref{eq:S2number}} & {4.7} \\  
 {Two-loop vertex} & {Last row of Table \ref{tab:V2}} & {16} \\  
 {Planar double box} & {Last row of Table \ref{tab:B2s}} & {71} \\    
 {Three-loop planar box} & {Eq.\eqref{eq:B3number}} & {22} \\ 
 {Two-loop pentabox} & {Eq.\eqref{eq:E2number}} & {17} \\
 {UV one-loop bubble} & {Last row of Table \ref{tab:BFDR}} & {0.17} \\ 
  {IR case}& {Table \ref{tab:KIR}} & {0.13} \\
\noalign{\smallskip}\hline
\end{tabular}
\end{table}

We have focused on scalar integrals without any structure in the numerator, but we expect that the treatment of loop tensors should follow the same guidelines described in this paper.  In particular, the approach of Sect.~\ref{sec:3}, in which an analytic integration is performed over the loop time-component, should work as it stands. As for the gluing method of Sect.~\ref{sec:4}, adding structures in the numerator could potentially lead to worse behaviour that needs to be corrected by local subtractions of large loop configurations, as done in Eqs. \eqref{eq:Csub}, \eqref{eq:Ssub}, \eqref{eq:BFDRgp}, {or by the technical cuts described in \eqref{eq:Lambdacut} and \eqref{eq:V2D}.} We leave a detailed study of this subject for further investigation.

We have sketched out how our method can be extended to UV and IR divergent configurations. Again, a deeper investigation is left for the future.

In summary, we believe that a numerical treatment of virtual corrections in four dimensions, of the type we have proposed, could be very beneficial in the computation of complicated multi-leg multi-scale amplitudes. 
{More specifically, we think that the direction to go would be to integrate directly the amplitude as a whole, rather than the separate MIs. This could mitigate some of the large gauge cancellations among individual contributions, if common loop momentum routings are chosen for classes of diagrams.}
In addition, Monte Carlo integration of the loops and over the phase-space of real emissions can be merged, potentially stabilising and speeding up the calculation.

\begin{acknowledgements}
The work of RP is supported by the SRA grant PID2019-106087GB-C21 (10.13039/501100011033), by the Junta de Andalucía grants A-FQM-467-UGR18 and P18-FR-4314 (FEDER), and by the COST Action CA16201 PARTICLEFACE.

\noindent The work of BW was partially supported by STFC HEP consolidated grants ST/P000681/1 and ST/T000694/1.
\end{acknowledgements}

\appendix

\section{Taylor expansions}
\label{app:a}
The Taylor expansions for integrals like \eqref{eq:C0} and \eqref{eq:D} can be obtained using the general result
\bqa\label{eq:Lmunu}
&&\dashint d^4q\frac{q^{\mu_1}\ldots q^{\mu_k} q^{\nu_1} \ldots q^{\nu_k}}
  {(q^2-m^2+i\eps)^n}\nl
&&\qquad = C_{n,k}\frac{i\pi^2}{m^{2n-2k-4}} g^{\{\mu_1\nu_1}\ldots g^{\mu_k\nu_k\}}
\eqa
where
\bqa\label{eq:Cnk}
 C_{n,k} =
 (-1)^n\,(-4)^k \frac{(n-k-3)!}{k!\,(n-1)!}\,,
 \eqa
which can be established by induction.

For \eqref{eq:C0}, we first make a shift of variable $q\to q+p_2$.  Then for $\tau_1= \tau_2=0$, $\mu_1= \mu_2=1$ we have
\bqa\label{eq:Csum}
C = \sum_{k,l=0}^\infty \dashint d^4q\frac{(2q\cdot p_1)^k(-2q\cdot p_2)^l}{(q^2-m^2+i\eps)^{k+l+3}}.
\eqa
Applying \eqref{eq:Lmunu}, and noting that terms with $l\neq k$ vanish as they would involve $\tau_1$ or $\tau_2$,
\bqa\label{eq:triangexp}
\frac{C}{i\pi^2} &=& \sum_{k=0}^\infty (-4)^k \frac{C_{2k+3,k}}{m^{2k+2}} p_{1}^{\mu_1}\ldots
p_{1}^{\mu_k} p_{2}^{\nu_1}\ldots  p_{2}^{\nu_k} \nl
&& \times g_{\{\mu_1\nu_1}\ldots g_{\mu_k\nu_k\}}\nl
&=&-\sum_{k=0}^\infty \frac{p_{1}^{\mu_1}\ldots
p_{1}^{\mu_k} p_{2}^{\nu_1}\ldots  p_{2}^{\nu_k}}{(2k+2)! \,m^{2k+2}}
g_{\{\mu_1\nu_1}\ldots g_{\mu_k\nu_k\}}\nl
&=&\frac{-1}{m^2}\sum_{k=0}^\infty\frac{(k!)^2}{(2k+2)!}\tau^k\,,
\eqa
where $\tau =2p_1\cdot p_2/m^2$.
For $\tau_1,\tau_2\neq 0$, we have instead of \eqref{eq:Csum}
\bqa
C &=& \sum_{k,l=0}^\infty \dashint d^4q\frac{(2q\cdot p_1-p_1^2)^k(-2q\cdot p_2-p_2^2)^l}{(q^2-m^2+i\eps)^{k+l+3}}\nl
&=&-\frac{i\pi^2}{2m^2}\Bigl[1 + \frac 1{12}(\tau+\tau_1+\tau_2)+\frac 1{90}(\tau^2+\tau_1^2+\tau_2^2\nl
&& +\tau\tau_1+\tau\tau_2+\tau_1\tau_2)+\ldots\Bigr]\,.
\eqa
This expansion can be extended to general masses by the substitutions
\bqa
\tau_1 &\to & \tau_1+\mu_2-\mu_1\,,\nl
\tau_2 &\to & \tau_2+\mu_2-1\,.
\eqa

For the expansion of \eqref{eq:D} we have
\bqa
\frac{D_0}{i\pi^2}
 &=&\sum_{j,k,l,n}  \frac{l!\,(-m^2\tau)^{l-n}}{n!(l-n)!}\nl
 &&\times\dashint d^4q
\frac{(-2p_1.q)^j (2p_2.q)^k [2(p_3-p_1).q]^n}
{(q^2-m^2+i\eps)^{j+k+l+4}}.
\eqa
Note that the number of factors of $q^\mu$ must be even, $j+k+n=2K$.
Then
\bqa
\frac{D_0}{i\pi^2}
&=&\sum_{j,k,l,n}(-1)^{j+l+n}\,4^{K}
\frac{l!\,\tau^{l-n}}{n!\,(l-n)!}
\frac{C_{N,K}}{m^{2K+4}}\nl
&&\times p_1^{\mu_1}\ldots p_1^{\mu_j}
p_2^{\mu_{j+1}}\ldots p_2^{\mu_{j+k}}\nl
&&\times(p_3-p_1)^{\mu_{j+k+1}}\ldots (p_3-p_1)^{\mu_{2K}}\nl
&&\times g_{\{\mu_1\mu_2}\ldots  g_{\mu_{2K-1}\mu_{2K}\}}
\eqa
where $N=j+k+l+4$. Substituting \eqref{eq:Cnk}, we obtain
\eqref{eq:tayD0}.

\section{The one-point FDR integral}
\label{app:b}
We compute the one-loop integral
\bqa
A_{\scriptscriptstyle \rm FDR} = \int [d^4q] \frac{1}{\bar D_0},
\eqa
with $\bar D_0$ given in Eq. (\ref{eq:den0}).
Extracting the {\em vacuum}  produces the expansion
\bqa
\frac{1}{\bar D_0}= 
     \left[\frac{1}{\bar q^2}\right]
+ \left[\frac{m^2_0}{\bar q^4}\right]
+ \frac{m^4_0}{\bar q^4\bar D_0},
\eqa
hence
\bqa
A_{\scriptscriptstyle \rm FDR} = \lim_{\mu \to 0}\left.\int d^4q
 \frac{m^4_0}{\bar q^4\bar D_0}\right|_{\mu = \mur}.
\eqa
By taking the ${\mu \to 0}$ limit at the integrand level and replacing $\mu$ with $\mur$ in the logarithmic divergent {\em vacuum} one obtains
\bqa
&&A_{\scriptscriptstyle \rm FDR} = \\
&&~\int d^4q
\left\{
 \frac{1}{q^2-m^2_0+i \epsilon}
-\frac{1}{q^2+i \epsilon}
-\frac{m^2_0}{(q^2-\mur^2+i \epsilon)^2}
\right\}.\nonumber
\eqa
Choosing now $m=m_0$ gives
\bqa
{A_{\scriptscriptstyle \rm FDR}}/{m^2}
= 4 \pi \int_0^{\infty}\! \rho^2 d\rho\, I_t,
\eqa
where
\bqa
I_t &:=& \int_{-\infty}^{+ \infty} dt
\left\{ 
  \frac{1}{t^2-R^2_0+i \epsilon}
- \frac{1}{t^2-\rho^2+i \epsilon} \right. \nl
&& \left.
- \frac{1}{(t^2-R^2_\nu+i \epsilon)^2}
\right\},
\eqa
with $R_0$ and $R_\nu$ defined in \eqref{eq:nuscale}.
One computes
\bqa
I_t = i \pi
\left(
\frac{1}{\rho}-\frac{1}{R_0}-\frac{1}{2 R^3_\nu}
\right),
\eqa
which gives
\bqa
\label{eq:AFDR}
{A_{\scriptscriptstyle \rm FDR}}/{m^2}=
4 i\pi^2 \dashint \frac{d \rho}{\rho}\,\theta(\rho) F(\rho),
\eqa
where
\bqa
F(\rho) := 
\frac{1}{\sqrt{1+\frac{1}{\rho^2}}
\left(1+\sqrt{1+\frac{1}{\rho^2}}\right)}
-
\frac{1}{2\left(1+\frac{\nu}{\rho^2}\right)^{\frac{3}{2}}}. \nonumber
\eqa
Note that there is no pole in this case.
In Table \ref{tab:AFDR} we report a comparison between a numerical implementation of \eqref{eq:AFDR} and the analytic result
\bqa
\label{eq:anala}
{A_{\scriptscriptstyle \rm FDR}}/{m^2}= i \pi^2 (1+\ln \nu).
\eqa

\begin{table}
\caption{The integral of \eqref{eq:AFDR} as a function of $\nu$ compared to the analytic result of \eqref{eq:anala}.
The numerical estimates are obtained with $10^8$ shots. MC errors between parentheses.}
\label{tab:AFDR}       
\begin{tabular}{rll}
\hline\noalign{\smallskip}
 $\nu$   & Numerical result & Analytic result \\
\noalign{\smallskip}\hline\noalign{\smallskip}
0.1 &   $-$$i$ 1.2855(2)$\times 10^{1}$ &  $-$$i$ 1.2856$\times 10^{1}$ \\
0.5 &      $i$ 3.0281(4) &     $i$ 3.0285 \\
1   &      $i$  9.869(1) &     $i$ 9.8696 \\
2   &      $i$ 1.6709(2)$\times 10^{1}$ &     $i$ 1.6711$\times 10^{1}$ \\
10  &      $i$ 3.2596(4)$\times 10^{1}$ &     $i$ 3.2595$\times 10^{1}$ \\
\noalign{\smallskip}\hline
\end{tabular}
\end{table}

\section{The IR integrals}
\label{app:c}
Here we obtain onefold integral representations for the cut-diagrams
${\rm D_{a,b,c,d}}$ of Sect.~\ref{sec:6} and 
prove \eqref{eq:IRuns}.
\paragraph{The diagram $\rm D_b$:}
$~$ \vskip 5pt
\noindent
Choosing the momenta as in \eqref{eq:routing1} gives
\bqa
\label{eq:Db0}
\frac{{\rm D_b}}{g^4} &=&\frac{8 \pi^3}{s} \int d^4 \omega_1 d^4 \omega_2
\frac{\delta_+(\sigma_2)\delta_+(\sigma_3-\eta)\delta_+(\sigma_4-\eta)}{(\sigma_1-\eta+i\epsilon)(\sigma_5-i \epsilon)} \nl
&&\times \theta(\rho_1+\rho_2 c_\theta),
\eqa
with
\bqa
\label{eq:sigmas}
\sigma_1 &=& (1-t_2)^2-\rho^2_2, \nl
\sigma_2 &=& (1-t_1-t_2)^2-\rho^2_1-\rho^2_2-2\rho_1 \rho_2 c_\theta,\nl
\sigma_3 &=& t^2_1-\rho^2_1,~~ \sigma_4 = t^2_2-\rho^2_2,\nl
\sigma_5 &=& \sigma_2-1+2(t_1+t_2).
\eqa
Note that a harmless $\mu^2$ has been added to propagator 1 and that the Heaviside function forces propagator 5 to have a positive component of the momentum along $x$. 
Using the three Dirac delta functions one arrives at
\bqa
\label{eq:Db2}
\frac{{\rm D_b}}{g^4} &=&-\frac{8 \pi^5}{s}\! \int_{\sqrt{\eta}}^{\infty}\! dR_1
\int_{\sqrt{\eta}}^{\infty}\! dR_2\, \frac{1}{1-2R_2} \frac{1}{1-2R^+} \nl
&& \times \theta(1-R^+) \theta(1-|c_\theta|) \theta(\rho_1+\rho_2 c_\theta),
\eqa
with $R_{1,2}$ in \eqref{eq:eta}, $R^+ := R_1+R_2$ and
\bqa
\label{eq:ctheta}
c_\theta= \frac{1}{2 \rho_1 \rho_2}
\Bigl[1+2(\eta+ R_1 R_2-R^+)
\Bigr].
\eqa
Integrating analytically over $R_2$ produces logarithms with boundaries determined by the three Heaviside functions. The result reads
\bqa
\label{eq:Db1}
\frac{\rm D_b}{g^4} &=&-\frac{2 \pi^5}{s} 
\int_{\sqrt{\eta}}^{\frac{1}{2}} \frac{dR_1}{R_1}
\ln \Bigg(\eta
\frac{1+\frac{1-2R_1}{R_1+\sqrt{R^2_1-\eta}}}{R_1+\sqrt{R^2_1-\eta}-\eta}
\Bigg). \nonumber \\
\eqa
\paragraph{The diagram $\rm D_a$:}
$~$ \vskip 5pt
\noindent We split $\rm D_a$ into two components, $\rm D_a=D^+_a+D^-_a$, with positive and negative values of $q_1$ along $x$.
Choosing the momenta as in \eqref{eq:routing1} produces
\bqa
\label{eq:Da0}
\frac{\rm D^+_a}{g^4} &=&\frac{4 i \pi^2 }{s} \int d^4 \omega_1 d^4 \omega_2
\prod_{j=1,3,4} \left(\frac{1}{\sigma_j-\eta+i \epsilon}
\right) \nl
&& \times \delta_+(\sigma_2)
\delta_+(\sigma_5) \theta(\rho_1+\rho_2 c_\theta),
\eqa
with the same $\sigma_{1\div5}$ of \eqref{eq:sigmas}.
Using the two delta functions gives
\bqa
\label{eq:Da+}
\frac{\rm D^+_a}{g^4} &=&\frac{8 i \pi^4}{s} \int_0^{\infty}  d \rho_1 \rho_1
\int_0^{\infty} d \rho_2 \rho_2\, I_1 \nl
&& \times 
\theta(1-|c_\theta|) \theta(\rho^2_1-\rho^2_2 +1/4),
\eqa
where
\bqa
c_\theta= \frac{1}{2 \rho_1 \rho_2}
\Bigl(\frac{1}{4}-\rho^2_1-\rho^2_2\Bigr)
\eqa
and
\bqa
I_1 &:=& \int_0^\infty dt_1 \bigg\{ \frac{1}{t^2_1-R^2_1+i \epsilon} 
\\
&& \times \frac{1}{(1/2+t_1)^2-R^2_2+i \epsilon} \frac{1}{(1/2-t_1)^2-R^2_2+i \epsilon} \bigg\}.\nonumber
\eqa
${\rm D^-_a}$ is obtained from \eqref{eq:Da+}
by replacing 
$$\theta(\rho^2_1-\rho^2_2 +1/4) \to
 \theta(-\rho^2_1+\rho^2_2 -1/4).$$ Hence
\bqa
\frac{\rm D_a}{g^4} &=&\frac{8 i \pi^4}{s} \int_0^{\infty}  d \rho_1 \rho_1
\int_0^{\infty} d \rho_2 \rho_2\, I_1 
\theta(1-|c_\theta|).
\eqa
Computing $I_1$ with the Cauchy integral theorem gives
\bqa
&&\frac{\rm D_a}{g^4} =-\frac{8 \pi^5}{s} \int_{\sqrt{\eta}}^\infty dR_1
\int_{\sqrt{\eta}}^\infty dR_2\, \theta(1-|c_\theta|) \\
&&\,\times
\bigg\{
\frac{1}{1+2R_2}
\frac{1}{1+2R^+}
-\frac{1}{1-2R_2+i \epsilon}
 \frac{1}{1-2R^++i \epsilon}
\bigg\}.\nonumber
\eqa
Note the appearance of the same denominator structures of \eqref{eq:Db2}.
An integration over $R_2$ produces
\bqa
\label{eq:Da1}
\frac{\rm D_a}{g^4} &=&-\frac{2 \pi^5}{s} \int_{\sqrt{\eta}}^\infty \frac{dR_1}{R_1}
\bigg\{\ln \frac{1+2R_2}{1+2R^+} \nl
&&-\ln \frac{1-2R_2+i\epsilon}{1-2R^++i\epsilon}
\bigg\}^{R^+_2}_{R^-_2},
\eqa
where
\bqa
R^\pm_2 := \sqrt{\bigg(\frac{1}{2}\pm\sqrt{R^2_1-\eta}\bigg)^2+\eta}.
\eqa
\paragraph{The diagram $\rm D_c$:}
$~$ \vskip 5pt
\noindent It is the complex conjugate of \eqref{eq:Da1}.
\paragraph{The diagram $\rm D_d$:}
$~$ \vskip 5pt
\noindent
Inserting a harmless $\mu^2$ in propagator 5 and
choosing the momenta as in \eqref{eq:routing2} gives
\bqa
\label{eq:Dd0}
\frac{{\rm D_d}}{g^4} &=&\frac{8 \pi^3}{s} \int d^4 \omega_1 d^4 \omega_2
\frac{\delta_+(\sigma_2-\eta)\delta_+(\sigma_3-\eta)\delta_+(\sigma_4)}{(\sigma_1+i\epsilon)(\sigma_5-\eta-i \epsilon)} \nl
&&\times \theta(c_\theta),
\eqa
where
\bqa
\sigma_1 &=& \sigma_4+1-2(t_2-t_1),
~~\sigma_2= (1-t_2)^2-\rho^2_2, \nl
\sigma_3 &=& t^2_1-\rho^2_1,~~
\sigma_4= (t_2-t_1)^2-\rho^2_1-\rho^2_2-2 \rho_1 \rho_2 c_\theta, \nl
\sigma_5 &=& t^2_2-\rho^2_2.
\eqa
Using the delta functions produces
\bqa
\label{eq:Dd2}
\frac{{\rm D_d}}{g^4} &=&-\frac{8 \pi^5}{s}\! \int_{\sqrt{\eta}}^{\infty}\! dR_1
\int_{\sqrt{\eta}}^{\infty}\! dR_2\, \frac{1}{1-2R_2} \frac{1}{1-2R^+} \nl
&& \times \theta(1-R^+) \theta(c_\theta) \theta(1-c_\theta),
\eqa
with $c_\theta$ as in \eqref{eq:ctheta}.
Integrating over $R_2$ gives
\bqa
\label{eq:Dd1}
\frac{\rm D_d}{g^4} &=&-\frac{2 \pi^5}{s} 
\int_{\sqrt{\eta}}^{R^+_1} \frac{dR_1}{R_1}
L(R_1),
\eqa
where
\bqa
L(R_1) &:=& \ln \Bigg[
\Bigg(\frac{1}{1-2 R_1}
    +\frac{1}{R_1+\sqrt{R^2_1-\eta}}
\Bigg) \\
&& \times
\frac{\eta (R_1-2\eta)}{(R_1+\sqrt{R^2_1-\eta}-\eta)(R_1(1-2 R_1)+2\eta)}
\Bigg], \nonumber
\eqa
and
\bqa
R^+_1 := \frac{1}{2}
\frac{1+4\eta^2}{1+\sqrt{\eta-4\eta^2(1-\eta)}}.
\eqa
\paragraph{The combination 
$\rm D^{u}_{a}+D^{u}_{c}$:}
$~$ \vskip 5pt
\noindent
This is obtained from \eqref{eq:Da1} and its complex conjugate by replacing the
integration region 
$\sqrt{\eta} < R_1 < \infty$ by
$1/2 < R_1 < \infty$. Upon doing so,
$\eta$ can be set to $0$ and
$R^\pm_2= R_1 \pm 1/2$.
Introducing $x= 2 R_1$ gives
\bqa
\frac{\rm D^{u}_{a}+D^{u}_{c}}{g^4}
= -4\frac{\pi^5}{s} \int_1^\infty \frac{dx}{x} 
\left(
\ln \frac{x+2}{x+1}+\ln \left| 
\frac{x-2}{x-1}
\right|
\right),\nonumber
\eqa
from which \eqref{eq:IRuns} follows.
\bibliographystyle{spphys}       
\bibliography{mcloops}   

\begin{thebibliography}{10}
\providecommand{\url}[1]{{#1}}
\providecommand{\urlprefix}{URL }
\expandafter\ifx\csname urlstyle\endcsname\relax
  \providecommand{\doi}[1]{DOI \discretionary{}{}{}#1}\else
  \providecommand{\doi}{DOI \discretionary{}{}{}\begingroup
  \urlstyle{rm}\Url}\fi

\bibitem{Heinrich:2020ybq}
G.~Heinrich, Phys. Rept. \textbf{922}, 1 (2021).
\newblock \doi{10.1016/j.physrep.2021.03.006}

\bibitem{Soper:1999xk}
D.E. Soper, Phys. Rev. D \textbf{62}, 014009 (2000).
\newblock \doi{10.1103/PhysRevD.62.014009}

\bibitem{Binoth:2000ps}
T.~Binoth, G.~Heinrich, Nucl. Phys. B \textbf{585}, 741 (2000).
\newblock \doi{10.1016/S0550-3213(00)00429-6}

\bibitem{Binoth:2005ff}
T.~Binoth, J.P. Guillet, G.~Heinrich, E.~Pilon, C.~Schubert, JHEP \textbf{10},
  015 (2005).
\newblock \doi{10.1088/1126-6708/2005/10/015}

\bibitem{Nagy:2006xy}
Z.~Nagy, D.E. Soper, Phys. Rev. D \textbf{74}, 093006 (2006).
\newblock \doi{10.1103/PhysRevD.74.093006}

\bibitem{deDoncker:2004fb}
E.~de~Doncker, Y.~Shimizu, J.~Fujimoto, F.~Yuasa, Comput. Phys. Commun.
  \textbf{159}, 145 (2004).
\newblock \doi{10.1016/j.cpc.2004.01.004}

\bibitem{Yuasa:2011ff}
F.~Yuasa, E.~de~Doncker, N.~Hamaguchi, T.~Ishikawa, K.~Kato, Y.~Kurihara,
  J.~Fujimoto, Y.~Shimizu, Comput. Phys. Commun. \textbf{183}, 2136 (2012).
\newblock \doi{10.1016/j.cpc.2012.05.018}

\bibitem{deDoncker:2017gnb}
E.~de~Doncker, F.~Yuasa, K.~Kato, T.~Ishikawa, J.~Kapenga, O.~Olagbemi, Comput.
  Phys. Commun. \textbf{224}, 164 (2018).
\newblock \doi{10.1016/j.cpc.2017.11.001}

\bibitem{Baglio:2020ini}
{Baglio, J. and Campanario, F. and Glaus, S. and M\"uhlleitner, M. and Ronca,
  J. and Spira, M. and Streicher, J.}, JHEP \textbf{04}, 181 (2020).
\newblock \doi{10.1007/JHEP04(2020)181}

\bibitem{Ghinculov:1996vd}
A.~Ghinculov, Phys. Lett. B \textbf{385}, 279 (1996).
\newblock \doi{10.1016/0370-2693(96)00871-4}

\bibitem{Guillet:2019hfo}
{Guillet, J. Ph. and Pilon, E. and Shimizu, Y. and Zidi, M. S.}, PTEP
  \textbf{2020}(4), 043B01 (2020).
\newblock \doi{10.1093/ptep/ptaa020}

\bibitem{Bauberger:2019heh}
{Bauberger, Stefan and Freitas, Ayres and Wiegand, Daniel}, JHEP \textbf{01},
  024 (2020).
\newblock \doi{10.1007/JHEP01(2020)024}

\bibitem{Pittau:2012zd}
R.~Pittau, JHEP \textbf{11}, 151 (2012).
\newblock \doi{10.1007/JHEP11(2012)151}

\bibitem{Kleiss:1994qy}
R.~Kleiss, R.~Pittau, Comput. Phys. Commun. \textbf{83}, 141 (1994).
\newblock \doi{10.1016/0010-4655(94)90043-4}

\bibitem{Kermanschah:2021wbk}
D.~Kermanschah, e-Print: 2110.06869 [hep-ph]  (2021)

\bibitem{vanHameren:2010cp}
A.~van Hameren, Comput. Phys. Commun. \textbf{182}, 2427 (2011).
\newblock \doi{10.1016/j.cpc.2011.06.011}

\bibitem{Broadhurst:1987ei}
D.J. Broadhurst, Z. Phys. C \textbf{47}, 115 (1990).
\newblock \doi{10.1007/BF01551921}

\bibitem{Ossola:2007ax}
G.~Ossola, C.G. Papadopoulos, R.~Pittau, JHEP \textbf{03}, 042 (2008).
\newblock \doi{10.1088/1126-6708/2008/03/042}

\bibitem{Pittau:2013qla}
R.~Pittau, Eur. Phys. J. C \textbf{74}(1), 2686 (2014).
\newblock \doi{10.1140/epjc/s10052-013-2686-1}

\bibitem{Gnendiger:2017pys}
C.~Gnendiger, et~al., Eur. Phys. J. C \textbf{77}(7), 471 (2017).
\newblock \doi{10.1140/epjc/s10052-017-5023-2}

\end{thebibliography}

\end{document}